\documentclass[twocolumn,showpacs,preprintnumbers,amsmath,amssymb,superscriptaddress,nofootinbib,english]{revtex4-1}
\usepackage{times,amsmath,amsfonts,amssymb,epstopdf}
\usepackage{graphicx}
\usepackage{dcolumn}
\usepackage{bm}
\usepackage{epsfig}
\usepackage{graphicx}
\usepackage{hyperref}
\usepackage[usenames]{color}
\usepackage{url}
\usepackage[normalem]{ulem}
\usepackage[T1]{fontenc}
\usepackage[dvipsnames]{xcolor}

\newcommand{\g}{\textcolor{black}}
\newcommand{\BLED}[1]{\textcolor{black}{#1}}
\newcommand{\HANS}[1]{\textcolor{black}{#1}}

\newcommand{\mpl}{M_{\rm Pl}}

\newcommand{\rd}{{\rm d}}

\def\eg{{\frenchspacing\it e.g.}}

\def\d{{\rm d}}
\def\be{\begin{equation}}
\def\ee{\end{equation}}
\def\ba{\begin{eqnarray}}
\def\ea{\end{eqnarray}}
\begin{document}

\title{Systematic simulations of modified gravity:  symmetron and dilaton models}

\author{Philippe~Brax}
\email[Email address: ]{philippe.brax@cea.fr}
\affiliation{Institut de Physique Theorique, CEA, IPhT, CNRS, URA 2306, F-91191Gif/Yvette Cedex, France}

\author{Anne-Christine~Davis}
\email[Email address: ]{a.c.davis@damtp.cam.ac.uk}
\affiliation{DAMTP, Centre for Mathematical Sciences, University of Cambridge, Wilberforce Road, Cambridge CB3 0WA, U.K.}

\author{Baojiu~Li}
\email[Email address: ]{baojiu.li@durham.ac.uk}
\affiliation{Institute for Computational Cosmology, Department of Physics, Durham University, Durham DH1 3LE, U.K.}

\author{Hans~A.~Winther}
\email[Email address: ]{h.a.winther@astro.uio.no}
\affiliation{Institute of Theoretical Astrophysics, University of Oslo, 0315 Oslo, Norway}

\author{Gong-Bo~Zhao}
\email[Email address: ]{gong-bo.zhao@port.ac.uk}
\affiliation{Institute of Cosmology \& Gravitation, University of Portsmouth, Portsmouth PO1 3FX, U.K.}
\affiliation{National Astronomy Observatories, Chinese Academy of Science, Beijing, 100012, P.R.China}

\date{\today}

\begin{abstract}
We study the linear and nonlinear structure formation in the dilaton and symmetron models of modified gravity using a generic parameterisation which describes  a large class of scenarios using only a few parameters, such as the coupling \g{between} the scalar field \g{and the matter,} and the range of the scalar force on very large scales. For this we have modified the $N$-body simulation code {\tt ECOSMOG}\g{, which is a variant of {\tt RAMSES} working in modified gravity scenarios,} to perform a set of $110$ simulations for different models and parameter values, including the default $\Lambda$CDM. These simulations enable us to explore a large portion of the parameter space. We have studied the effects of modified gravity on the matter power spectrum and mass function, and found a rich and interesting phenomenology where the difference with the $\Lambda$CDM template cannot be reproduced by a linear analysis even on scales as large as $k\sim0.05$ $h\rm{Mpc}^{-1}$.
Our results show the full effect of screening on nonlinear structure formation and the associated  deviation \g{from} $\Lambda$CDM. \BLED{We also investigate how differences in the \g{force mediated by the scalar field in} modified gravity models lead to qualitatively different features for the nonlinear power spectrum and the halo mass function, and how varying the individual model parameters changes these observables. The differences are particularly large in the nonlinear power spectra whose shapes for $f(R)$, dilaton and symmetron models vary greatly, and where the characteristic bump around $1\ h\rm{Mpc}^{-1}$ of $f(R)$ models is preserved for symmetrons, whereas an  increase on much smaller scales is particular to symmetrons. No bump is present for dilatons where a flattening of the power spectrum takes place on small scales.   These deviations from \g{$\Lambda$CDM} and the differences between modified gravity models, such as dilatons and symmetrons,  could be tested with future surveys.  }
\end{abstract}


\maketitle

\section{Introduction}

\label{sect:introduction}

The apparent acceleration of the Universe could be due to at least four different reasons: a cosmological constant, dark energy \citep{cst2006}, modified gravity \citep{cfps2011} or large spatial inhomogeneities \cite{biswas}. The last of these violates the Copernican principle and requires a theory for the initial conditions of the Universe  while the first three invoke a change of the dynamics of the Universe itself.

The cosmological constant solution is rather peculiar as no real dynamics is attached to it until the vacuum energy starts dominating the energy content of the Universe. This seems to have happened in the quite recent past, a fact which is problematic and related to the astoundingly  small value of the critical density of the Universe compared to particle physics expectations, which scale as the fourth power of the mass of any heavy particle present in the early Universe.

To alleviate this problem, two other possibilities are commonly invoked. The first one is dark energy \citep{cst2006}, in which the dynamics of a field \g{(\eg, a scalar field in the simplest case)} determines the \g{fate of the Universe}. So far no real solution to the cosmological constant problem has been found within this setting although phenomenological works abound. Setting aside the problem of the actual value of the dark energy density now, these models suffer from another serious problem: \HANS{dark energy evolves on cosmological time scales only when the scalar field leads to a long range interaction.} Of course, one can decree that dark energy does not couple to baryons as in coupled quintessence models\footnote{\BLED{We regard the coupled quintessence model as an example of dark energy rather than modified gravity, for which we require a universal coupling to all matter species.}}, and therefore alleviate gravitational problems linked to the existence of a scalar fifth force. If this is not the case, then a solution which has been put forward in the last decade is screened modified gravity mediated by a scalar field.


Many models of \BLED{screened} modified gravity have been constructed so far, which fall within two broad categories. Following the initial works on massive gravity, models involving nonlinear kinetic terms, such as the Galileon \citep{dgp2000,nrt2009,dev2009}, make use of the Vainshtein mechanism \cite{vainshtein}  whereby large nonlinearities in the vicinity of dense objects effectively reduce the scalar coupling to matter to be below the experimental bounds. Another class of models originating from the chameleon theory \citep{kw2004,ms2007} use a screening of the fifth force in dense environments due to the nonlinearities of either the scalar potential or its coupling to matter (or both). Chameleon models such as $f(R)$ gravity \citep{lb2007,hs2007,bbds2008} are such that the mass of the scalar field becomes large in dense bodies, effectively suppressing the magnitude of the scalar force; other models such as the dilatons \citep{bbds2010} and symmetrons \citep{hk2010,pospelov} are such that the effective coupling to matter becomes vanishingly small in dense environments. All cases in the second class of screened modified gravity can be described by the same formalism which has been recently unified \citep{bdl2011,bdlw2012}. In this paper, we will concentrate on the second class.

It has been shown in \citep{bdlw2012} that the background cosmology of these models is extremely constrained. Indeed, the fact that particle masses (in the Einstein frame) and the gravitational constant (in the Jordan frame) cannot vary substantially between \g{the era of the} Big Bang Nucleosynthesis (BBN) and now implies that the scalar field must stay very close to the minimum of the effective potential since before BBN. This is guaranteed when the mass of the scalar field on the cosmological background is much heavier than the Hubble expansion rate, securing the stability of the minimum to `kicks' occurring when particles such as the electron\g{s} decouple \citep{bbdkw2004}. A consequence of this is that the effective equation of state of the scalar field in the late-time Universe becomes extremely close to $-1$, hardly distinguishable from the pure $\Lambda$-cold dark matter ($\Lambda$CDM) scenario. In practice, models of $f(R)$ gravity, chameleon, dilaton and symmetron types usually behave like $\Lambda$CDM in \HANS{the} background cosmology since before BBN.

Fortunately, this does not imply that their cosmology is totally \g{degenerate with that of the $\Lambda$CDM model}: the effects of modified gravity appear \g{in the structure formation}. Indeed, within the Compton wavelength of the scalar field\footnote{\BLED{The Compton wavelength of a scalar field is defined as $\lambda\equiv m^{-1}_{\rm eff}$, and $m_{\rm eff}$ is the effective mass of the scalar field (see below).}}, gravity is modified and the growth rate of structures is altered \citep{bbdkw2004,bdlw2012}. At the linear level, this results in a modification of the growth equation which depends on the scalar field mass $m(a)$ and the coupling to matter $\beta (a)$ expressed as functions of the scale factor. It turns out that all screened modified gravity models \BLED{ with no higher derivative terms in their Lagrangian, including their \g{field-dependent} potential $V(\varphi)$ and the coupling to matter $\beta(\varphi)$}, can be {\it fully} reconstructed from the {\it sole} knowledge of the functions $m(a)$ and $\beta (a)$. This allows one to engineer models directly from their linear perturbation properties, i.e., given $m(a)$ and $\beta(a)$ one can build a fully consistent model of modified gravity defined by $\beta(\varphi)$ and $V(\varphi)$ \citep{bdl2011,bdlw2012}, which implies that one could study the nonlinear evolution of cosmic structures in the late Universe simply from the knowledge of $m(a)$ and $\beta (a)$. This provides a {\it systematic} approach to screened modified gravity which can be applied to generalised chameleon, dilaton and symmetron models. For other schemes to parameterise modified gravity see \citep{ccm2007,aks2008, jz2008,s2009,fs2010,bfsz2011}.

Studying the nonlinear regime of structure formation is of particular importance for screened modified gravity models, as local gravity tests often imply that deviations from general relativity are strongest on megaparsec (Mpc) scales \citep{bdlw2012}, where nonlinearities cannot be neglected. Two competing effects influence the dynamics of modified gravity here. On the one hand, the gravitational interaction is enhanced by the presence of a long-range fifth force which implies an increase of the growth of structure. On the other hand, where local matter densities are high enough, screening effects develop and structure formation converges to its GR behaviour. These two competing effects have been confirmed in already-available $N$-body simulations of $f(R)$ gravity \citep{o2008,olh2008,sloh2009,zlk2011,zlk2011b,lh2011,lzk2012,lzlk2012,jblzk2012,lhkzjb2012}, chameleon \citep{lz2009,zmlhf2010,lz2010,li2011}, dilaton \citep{bbdls2011} and symmetron \citep{dlmw2012,wml2012} models.

In this work, we apply the $(m(a), \beta(a))$ parameterisation to generalise dilaton and symmetron models and study their large-scale structure formation. We use modified versions of the {\tt ECOSMOG} code \citep{ecosmog} to run $N$-body simulations in these models. This code is based on the publicly-available adaptive mesh refinement (AMR) code {\tt RAMSES} \citep{ramses}, which is efficiently parallelised and suitable to run simulations systematically. The AMR nature of the code means that a higher  resolution can be achieved, without sacrificing the overall performance of the code, in dense regions where the field equations are most nonlinear, ensuring the accuracy of the fifth force calculation there. As a result, our simulations are able to probe the structure formation in these modified gravity models down to scales well below the typical dark matter halo sizes.

The results of our simulations indicate that large deviations from $\Lambda$CDM in the power spectrum can be found on scales of order 1 Mpc for both symmetron and dilaton models for values of the parameters which comply with the local constraints $($the gravitational tests in the Solar system and a mild suppression of the fifth force on galactic scales typically impose that the range of the fifth force should be less than a few Mpc in the cosmological background$)$. Large differences are also present in the number density of intermediate-sized dark matter halos with masses of order $10^{13}-10^{14}h^{-1} M_\odot$ \HANS{(representing objects from groups of galaxies to small galaxy clusters)}.  For models with a fifth force whose range in the cosmological background is of order Mpc and a coupling strength to matter of order unity, the deviation from $\Lambda$CDM can reach $\sim40$\% in the symmetron case and $\sim30$\% in the dilatonic one. Such large differences are testable using future galaxy surveys.

Moreover, symmetron and dilaton models are distinguishable thanks to the very different time dependence of their couplings to matter. For symmetrons, the coupling has a slow dependence on the scale factor $a$ in the recent past of the Universe and vanishes before a transition redshift $z_\ast$ (its definition will be given later). Dilaton models have a much sharper dependence on the scale factor and generically decrease exponentially fast going back in time. \BLED{As will be discussed in detail in \S~\ref{subsect:tomography}, the time dependence of the coupling strength can be roughly translated into a density dependence, and the steep density dependence in the recent past of the Universe (or equivalently in regions of low matter density) for dilaton models suggests that the dilaton screening is more efficient}. These properties make the matter power spectra and halo mass functions behave qualitatively differently in these models. \BLED{We will give a more detailed summary of the results in the concluding section.}

The layout of this paper is as follows: in \S~\ref{sect:mod_grav} we review scalar-tensor theories and show how such theories of modified gravity can be analysed using a simple parametrisation which encapsulates all the dynamics; in \S~\ref{sect:models} we briefly describe the generalised symmetron (\S~\ref{subsect:symmetron_model}) and dilaton (\S~\ref{subsect:dilaton_model}) models and the possible effects of varying each model parameter; the equations that will be used in the $N$-body simulations are summarised in \S~\ref{sect:nbody_eqns}, while the details are given in \S~\ref{sect:discrete_eqns}; we next carry out tests of our codes in \S~\ref{sect:code_tests}, and the cosmological simulations of this work are then discussed in \S~\ref{sect:sim} for the symmetron (\S~\ref{subsect:sim_symmetron}) and dilaton (\S~\ref{subsect:sim_dilaton}) cases respectively; finally we summarise and conclude in \S~\ref{sect:summary}.

\BLED{In the paper we use the \HANS{units $\hbar = c=1$} except  where $c$ appears explicitly. Overbar (subscript $_0$) denotes the background (present-day) value of a quantity and subscript $_\varphi$ means ${\rm d}/{\rm d}\varphi$. $\kappa=8\pi G_N=M_{\rm Pl}^{-2}$, where $M_{\rm Pl}$ is the reduced Planck mass and $G_N$ is Newton's constant, are used interchangeably.} 
\\\\

\section{Modifying Gravity with a Scalar Field}
\label{sect:mod_grav}

\subsection{Screened modified gravity}

The action governing the dynamics of a scalar field $\varphi$ in a scalar-tensor theory is
of the general form
\begin{eqnarray}\label{eq:action}
S &=& \int {\rm d}^4x\sqrt{-g}\left[\frac{M_{\rm Pl}^2}{2}{R}-\frac{1}{2}(\nabla\varphi)^2- V(\varphi)\right]\nonumber\\
&& + \int {\rm d}^4x \sqrt{-\tilde g} {\cal L}_m(\psi_m^{(i)},\tilde g_{\mu\nu}),
\end{eqnarray}
where $g$ is the determinant of the metric $g_{\mu\nu}$, ${ R}$ is the Ricci scalar and $\psi_m^{(i)}$ are various matter fields labelled by $i$. A key ingredient of the model is the conformal coupling of $\varphi$ with matter particles. More precisely, the excitations of each matter field $\psi_m^{(i)}$ couple to a metric $\tilde g_{\mu\nu}$ which is related to the Einstein-frame metric $g_{\mu\nu}$ by the conformal rescaling
\be
\tilde g_{\mu\nu}=A^2(\varphi)g_{\mu\nu}.
\ee
The metric $\tilde g_{\mu\nu}$ is the Jordan-frame metric.
The fact that the scalar field couples to matter implies that the scalar field equation becomes density-dependent. More specifically, the scalar field equation of motion (EOM) is modified due to the coupling of the scalar field $\varphi$ to matter:
\be
\Box \varphi= -\beta T + \frac{{\rm d}V}{{\rm d}\varphi},
\ee
where $T$ is the trace of the energy momentum tensor $T^{\mu\nu}$, $\Box\equiv\nabla^\mu\nabla_\mu$ and the coupling of $\varphi$ to matter is defined by
\be
\beta(\varphi) \equiv \mpl\frac{{\rm d}\ln A}{{\rm d} \varphi}.
\ee
This is equivalent to the usual scalar field EOM with the effective potential
\be
V_{\rm eff}(\varphi) = V(\varphi) - \left[A(\varphi)-1\right]T.
\ee
We will always require that the effective potential possesses a unique density-dependent minimum in the presence of pressureless matter for which $T=-\rho_m$, i.e., that the potential
\be
V_{\rm eff}(\varphi) = V(\varphi) +[A(\varphi)-1]\rho_m
\ee
has a minimum $\varphi_{\rm min}(\rho_m)$. The mass of the scalar field at the minimum,
\be
m^2 = \frac{\rd^2 V_{\rm eff}}{\rd\varphi^2}\big|_{\varphi_{\rm min}},
\ee
must be positive.  In a cosmological setting we will also impose that $m^2\gg H^2$ with $H$ being the Hubble expansion rate. This guarantees the stability of the minimum to perturbations.

When matter is described by a pressure-less fluid with
\be
T^{\mu\nu}= \rho_m u^\mu u^\nu,
\ee
where $u^\mu\equiv\rd x^\mu/\rd\tau$ is the 4-velocity field of the fluid and $\tau$ is the proper time, the matter density $\rho_m$ is conserved
\be
\dot \rho_m +\theta\rho_m=0,
\ee
where $\theta\equiv\nabla_\mu u^\mu=3H$ is the expansion scalar and the trajectories are determined by the modified geodesics
\be
\dot u^\mu + \beta \frac{\dot \varphi}{M_{\rm Pl}} u^\mu= - \beta \frac{\nabla^\mu \varphi}{M_{\rm Pl}}.
\label{new}
\ee
In the weak-field limit with a line element
\be
\rd s^2=-(1+2\phi) \rd t^2+ (1-2\phi)\rd x^i\rd x_i,
\ee
and in the non-relativistic case, this reduces to the modified geodesic equation for matter particles
\be
\frac{\rd^2 x^i}{\rd t^2}= -\nabla^i\left[\phi+\ln A(\varphi) \right].
\ee
This can be interpreted as the motion of a particle in the effective gravitational potential defined as
\be
\Psi\equiv\phi+\ln A(\varphi),
\ee
and is a manifestation of the dynamics of modified gravity. \BLED{One may also  call the deviation from the Newtonian gravity a fifth force. In this paper we will use these terminologies interchangeably.}


When a particle of mass $M$ in a homogeneous background matter density is the source of gravity, the scalar field satisfies
\be\label{eq:kleqn}
\left(\nabla^2+m^2\right) \varphi= \beta \frac{M}{\mpl}\delta^{(3)}(r),
\ee
in which $\delta^{(3)}(r)$ is the 3-dimensional Dirac $\delta$-function and $m$ the scalar field mass in the background. This implies that
\be\label{eq:klsol}
\Psi= -\left(1+2\beta^2e^{-mr}\right)\frac{G_N M}{r}.
\ee
When $\beta\sim{\cal O}(1)$ and \BLED{ $mr\lesssim1$}, this implies a substantial deviation from Newton's law. For bodies much bigger than a point particle, 
nonlinear effects imply that the effective coupling felt by \BLED{ a test mass near the source} can be much smaller than $1$ or the \BLED{scalar field} mass becomes much larger than the inverse of the typical size of the \BLED{ source} ($m^{-1}\ll r$). The dilaton and symmetron models satisfy the first criterion which guarantees that solar system and laboratory tests of gravity are evaded.

\BLED{In addition to the {\it self}-screening described above, the modification of gravity depends on the environment of the bodies as well. For example, in a high-density background, the scalar field mass $m$ in Eq.~(\ref{eq:kleqn}) can be very large, which suppresses the deviation from Newtonian gravity according to Eq.~(\ref{eq:klsol}).}

This environmental dependence is at the heart of the  screening mechanisms in chameleon, dilaton and symmetron cases. Indeed, as shown in \citep{bdlw2012}, the screening is effective when the Newtonian potential $\Phi_N$ generated at the surface of a dense body satisfies
\begin{equation}\label{eq:sc}
\vert \varphi_\infty -\varphi_c\vert \ll 2 \beta_\infty M_{\rm Pl} \Phi_N,
\end{equation}
where $\varphi_{c,\infty}$ are respectively the minimum of the effective potential inside and far away from the dense body; $\Phi_N$ is the Newton potential at the surface of the body and $\beta_\infty =\beta (\varphi_\infty)$  is the coupling to matter outside. \BLED{Note that the self and environmental screenings are encoded in $\Phi_{N}$ and $\varphi_{\infty}, \beta_{\infty}$ respectively}

In cosmological simulations, $\varphi_\infty=\bar{\varphi}$ is the background value of $\varphi$, while $\varphi_c$ is the value inside clustered structures, which can be very small. In general, $\varphi$ could change \HANS{by several} orders of magnitude from low-density to high-density regions, and this is why the accurate calculation of $\varphi$ is a challenging task. The equations of motion which govern the dynamics of the modified gravity models which we consider here are
\begin{eqnarray}
\label{eq:newton}\nabla^2\phi &\approx& 4\pi G\left(\rho_m-\bar{\rho}_m\right),\\
\label{eq:sf}c^2\nabla^2\varphi &\approx& V_\varphi(\varphi)-V_\varphi(\bar{\varphi})+A_\varphi(\varphi)\rho_m-A_\varphi(\bar{\varphi})\bar{\rho}_m,\\
\label{eq:particle}\frac{{\rm d}^2\vec{r}}{\rd t^2} &=& -\vec{\nabla}\phi - c^2\BLED{\beta(\varphi)}\vec{\nabla}\varphi-\BLED{\beta(\varphi)}\dot{\varphi}\frac{\rd \vec{r}}{\rd t},
\end{eqnarray}
where in Eq.~({\ref{eq:newton}--\ref{eq:sf}) we have worked in the quasi-static limit so that terms involving time derivatives \HANS{have been} dropped; this is a good approximation throughout the course of cosmic evolution as the time derivatives are generally much smaller than the spatial ones\footnote{\BLED{This has been shown explicitly in, e.g., \cite{o2008}, which compares the two directly. A more rigorous proof of the validity of the quasi-static approximation would be by solving the full time-dependent scalar field EOM, which is beyond the scope of the current work. However we find that, in the linear perturbation calculations of \citep{bdlw2012}, one gets indistinguishable results by solving the full (linearised) EOM and using the quasi-static approximation, showing that the latter is actually quite reasonable.}}. The first of these equations is the Poisson equation while the last one is the modified Newtonian dynamics due to the presence of the scalar field $\varphi$, c.f.~Eq.~(\ref{new}). We have reinstated the factors of $c$ \BLED{because in code units (see below) $c$ is no longer unity}.

\subsection{Tomography}

\label{subsect:tomography}

We shall always consider the cosmological evolution of the scalar field $\varphi$ in modified gravity models with a minimum of $V_{\rm eff}(\varphi)$ at which the scalar field mass $m$ satisfies $m^2 \gg H^2$. The time evolution of the scalar field is tightly constrained by BBN physics due to its coupling to matter particles. The fact that the scalar field evolves along the minimum of $V_{\rm eff}(\varphi)$ implies that the masses of fundamental particles
\be
m_\psi= A(\varphi) m_{\rm bare},
\ee
in  which $m_{\rm bare}$ is the bare mass appearing in the matter Lagrangian, evolve too. In practice, tight constraints on the time variation of masses since the time of BBN
\be
\frac{\Delta m_\psi}{m_\psi}= \beta \frac{\Delta \varphi}{\mpl},
\ee
where $\Delta \varphi$ is the total variation of the field since BBN, impose that $\Delta m_\psi/m_\psi$ must be less than $\sim10\%$. At a redshift of order $z_e\approx 10^9$, electrons decouple and give a `kick' \cite{bbdkw2004}  to the scalar field which would lead to a large violation of the BBN bound. To avoid this, the field must be close to the minimum of $V_{\rm eff}(\varphi)$ before $z_e$ and simply  follow the time evolution of the minimum. Moreover, the total excursion of the scalar field following the minimum must be small enough. In practice, we will always assume that $|\varphi/\mpl|\ll 1$ along the minimum trajectory, implying that the BBN bound for the time dependent minimum is always satisfied. The models are then valid provided the electron `kick' does not perturb the minimum too much.
The minimum of the effective potential acts as a slowly varying cosmological constant. Indeed, when $m^2 \gg H^2$  the minimum is stable for all the models we will consider. In this case, the dynamics are completely determined by the minimum equation
\be
\frac{\rd V}{\rd\varphi}\Big\vert_{\varphi_{\rm min}}= -\beta A \frac{\rho_m}{\mpl}.
\end{equation}

In fact, the knowledge of the time evolution of the mass $m$ and the coupling $\beta$ is enough to determine the time evolution of the field.
Using the minimum equation, we can deduce that the field evolves according to
\begin{eqnarray}\label{eq:min_eqn}
\frac{\rd\varphi}{\rd t}=\frac{3H}{m^2} \beta A \frac{\rho_m}{\mpl}.
\end{eqnarray}
This is the time evolution of the scalar field at the background level since the instant when the field starts being at the minimum of the effective potential.
The knowledge of the time evolution of the mass $m$ and the coupling $\beta$ is enough to determine the bare potential $V(\varphi)$ and the coupling function $A(\varphi)$ completely. To see this, integrating Eq.~(\ref{eq:min_eqn}) once, we find
\begin{equation}\label{eq:Vofphi}
\varphi(a)=  \frac{3}{\mpl}\int_{a_{\rm ini}}^a \frac{\beta (a)}{a m^2(a)}\rho_m(a)\rd a +\varphi_c,
\end{equation}
where $\varphi_c$ is the initial value of the scalar field at $a_{\rm ini}<a_{\rm BBN}$ and we have taken $A(\varphi)\approx1$ given that the temporal variation of fermion masses must be very weak. If the coupling strength $\beta$ is expressed in terms of the field $\varphi$ and not the scale factor $a$, this is also equivalent to
\be
\int_{\varphi_c}^\varphi \frac{\rd\varphi}{\beta(\varphi)}=  \frac{3}{\mpl}\int_{a_{\rm ini}}^a \frac{1}{a m^2(a)}\rho_m(a)\rd a.
\ee
Similarly the minimum equation implies that the potential can be reconstructed as a function of time
\begin{equation}\label{eq:V}
V=V_0 - \frac{3}{\mpl^2}\int_{a_{\rm ini}}^a \frac{\beta^2(a)}{am^2(a)} \rho_m^2(a) \rd a,
\end{equation}
where $V_0$ is the value of the potential at $a=a_{\rm ini}$. This defines the bare scalar field potential $V(\varphi)$ parametrically when $\beta (a)$ and $m(a)$ are given. Hence we have found that the {\it full} nonlinear dynamics of the theory can be recovered from the knowledge of the {\it time} evolutions of the mass and the coupling to matter since before BBN.

The reconstruction mapping gives a one-to-one correspondence between the scale factor $a$ and the value of the field $\varphi(a)$ in the cosmic background. As the scale factor is in a one-to-one correspondence with the matter energy density $\BLED{\bar{\rho}_m(a)}$, we have obtained a mapping $\rho_m \to \varphi(\rho_m)$  defined using the time evolution of $m(a)$ and $\beta(a)$ only. Given these evolutions, one can reconstruct\footnote{\BLED{This is done by assuming that the scalar field always minimises its effective potential $V_{\rm eff}$, and thus the results below are more of qualitative estimates than quantitatively accurate predictions.}} the dynamics of the scalar field for densities ranging from cosmological to solar system values using Eq.~(\ref{eq:Vofphi}) and Eq.~(\ref{eq:V}). By the same token, $V(\varphi)$ can be reconstructed for all values of $\varphi$ (and $\rho_m$) of interest, from the solar system and Earth to the cosmological background today.

In particular, we can now state the screening condition of modified gravity models [c.f.~Eq.~(\ref{eq:sc})] as
\be\label{eqq}
\int_{a_{\rm in}}^{a_{\rm out}} \frac{\beta (a)}{a m^2(a)}\rho_m(a)\rd a \ll \beta_{\rm out}\mpl^2\Phi_N,
\ee
with constant matter densities $\rho_{\rm in,out}= \rho_m(a=a_{\rm in,out})$ inside and outside the dense body respectively, and where we have defined $\beta_{\rm out}\equiv\beta(a=a_{\rm out})$. Note that the gravitational properties of the screened modified gravity models can be captured by the cosmological evolutions of the scalar field mass and coupling function only.


The loosest screening condition follows from the fact \BLED{ the Milky Way should be screened as otherwise large deviations from Newtonian gravity would have been detected in the solar system}. For the Milky Way, the density is around six orders of magnitude larger than the cosmological background implying that $a_{\rm in}\sim 10^{-2}$; its Newtonian potential is $\Phi_G \sim 10^{-6}$. Taking the outside environment to be close to the cosmological background we have $a_{\rm out}\sim 1$. Writing
\be
m(a) = m_0 f(a), \ \ \ \ \beta (a)= \beta_0 g(a),
\ee
where $f$ and $g$ are smooth functions of $a$ with slow variations we find
\be
\frac{3\Omega_{m0}H_0^2}{m_0^2}\int_{a_{\rm in}}^{1} \frac{g(a)}{a^4 f^2(a)}\rd a \le \mpl^2\Phi_G,
\ee
\BLED{in which $\Omega_m$ is the fractional matter density}. Defining $I\equiv\int_{a_{\rm in}}^{1} \frac{g(a)}{a^4 f^2(a)}\rd a $, we find that
\be
\frac{m_0^2}{H_0^2}\ge \frac{3\Omega_{m0} I}{\Phi_G}.
\ee
Typically this implies that $m_0/H_0 \gtrsim 10^3$. Hence we find that screened models of modified gravity can only act on scales below the order of a few Mpc.
In fact we will make use of the ratio
\be
\xi\equiv\frac{H_0}{m_0},
\ee
which is related to the range of the fifth force as
\be
\lambda=2998\xi~h^{-1}{\rm Mpc}.
\ee
These scales, in the Mpc range, are beyond the linear perturbation regime and can only be accurately analysed using numerical simulations. This is the aim of the present article.
In the next subsection, we will describe the models we will study in detail numerically.

\subsection{The dilaton and symmetron models}

\subsubsection{Dilatons}

The environment-dependent dilaton model was originally described in \cite{bbds2010}. The essential features of the dilaton model include a runaway potential and
a coupling function $A(\varphi)$ which has a minimum. The potential is derived
in the strong coupling limit of string theory and the form of the coupling
function ensures the field does not runaway to infinity, which would imply
decompactification.
In \citep{bbds2010} the coupling function and bare potential of the scalar field were specified as follows:
\begin{eqnarray}
A(\varphi) &=& 1 + \frac{1}{2}\frac{A_2}{M^2_{\rm Pl}}\left(\varphi-\varphi_\ast\right)^2,\\
V(\varphi) &=& V_0e^{-\gamma\varphi/M_{\rm Pl}}.
\end{eqnarray}
Here $A_2\gg1, \gamma>0$ are dimensionless model parameters, $V_0$ is a model parameter with mass dimension 4 and $\varphi_\ast$ an arbitrary constant. The screening mechanism of the dilaton model is shown in Fig.~\ref{fig:dilaton}. Again, denoting the value of $\varphi$ which minimises $V_{\rm eff}(\varphi)$ by $\varphi_{\rm min}$, when matter density is high $\varphi_{\rm min}$ is very close to $\varphi_\ast$ so that $\beta(\varphi_{\rm min})\approx\beta(\varphi_\ast)=0$ and the fifth force essentially vanishes, while when matter density is low $\varphi_{\rm min}$ can evolve away from $\varphi_\ast$ so that $\beta(\varphi_{\rm min})\neq\beta(\varphi_\ast)=0$, giving rise to a non-negligible fifth force.

\begin{figure*}
\includegraphics[scale=0.5]{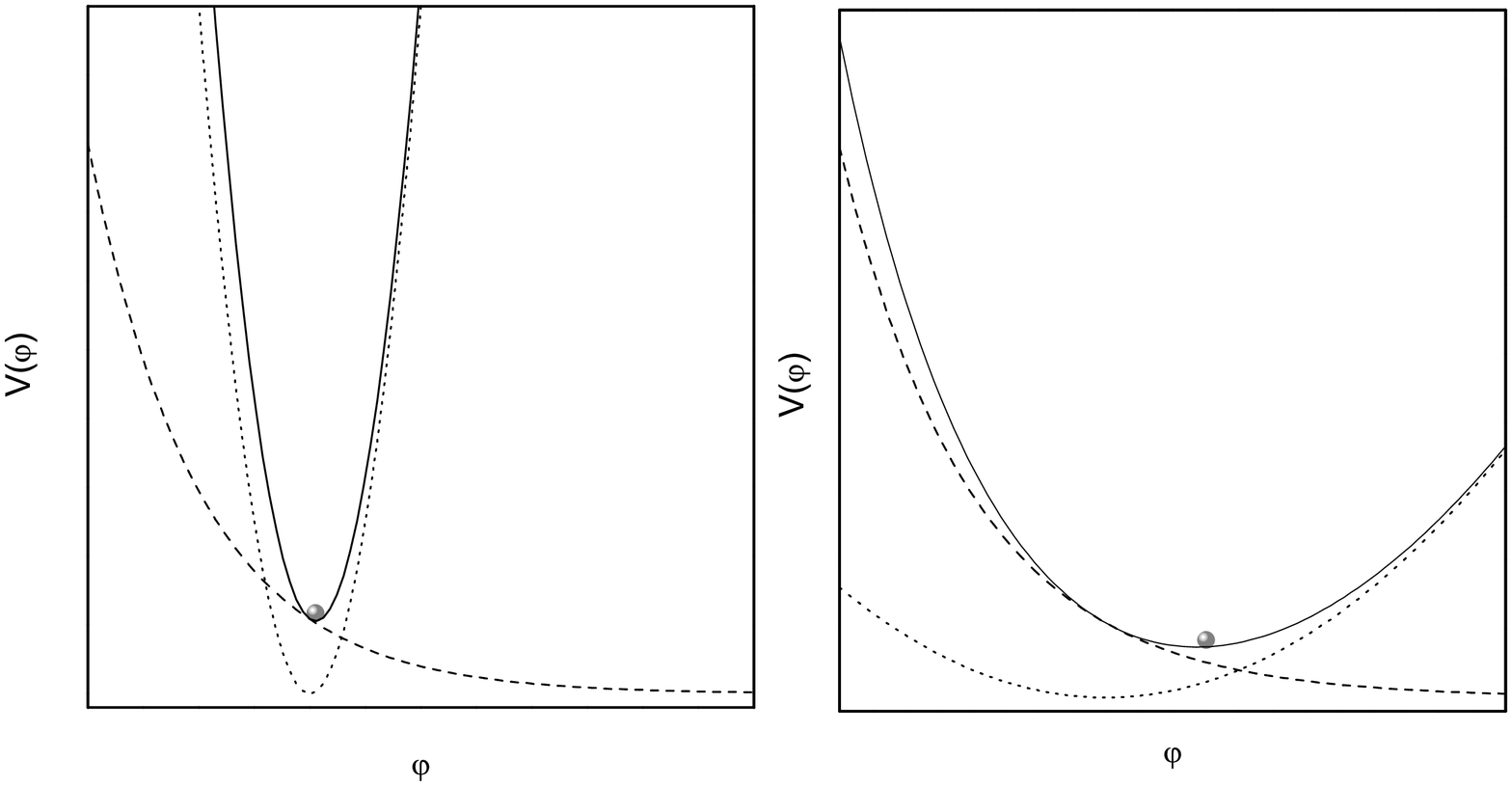}
\caption{An illustration of how the dilaton mechanism works. The dashed, dotted and solid curves are respectively the bare potential $V(\varphi)$ of the dilaton field, the coupling function and the total effective potential $V_{\rm eff}(\varphi)$. {\it Left Panel}: in high matter-density regions the minimum of $V_{\rm eff}(\varphi)$ is where the coupling strength vanishes and so the fifth force is suppressed. {\it Right Panel}: in low matter-density regions the coupling strength does not vanish at the minima of $V_{\rm eff}(\varphi)$, where the dilaton field resides, and so a nonzero fifth force takes effect in structure formation.} \label{fig:dilaton}
\end{figure*}

To study the
cosmology of the dilaton model we need only consider the dynamics in the
vicinity of the field $\varphi_ \ast$, where
\be
\beta (\varphi) \approx \frac{A_2}{\mpl}(\varphi-\varphi_ \ast),
\ee
from which we deduce that
\be
\ln\left\vert\frac{\varphi-\varphi_ \ast}{\varphi_c-\varphi_ \ast}\right\vert = 9 A_2\Omega_{m0}H_0^2 \int_{a_{\rm ini}}^a \frac{\rd a}{a^4 m^2 (a)},
\ee
and therefore
\begin{equation}
\vert \beta (\varphi)\vert = \vert \beta (\varphi_c)\vert  \exp\left[9 A_2\Omega_{m0}H_0^2 \int_{a_{\rm ini}}^a \frac{\rd a}{a^4 m^2 (a)}\right].
\label{eq:beta}
\end{equation}
This is the relation between the coupling at the initial time and other cosmological times.

\begin{figure*}
\includegraphics[scale=0.5]{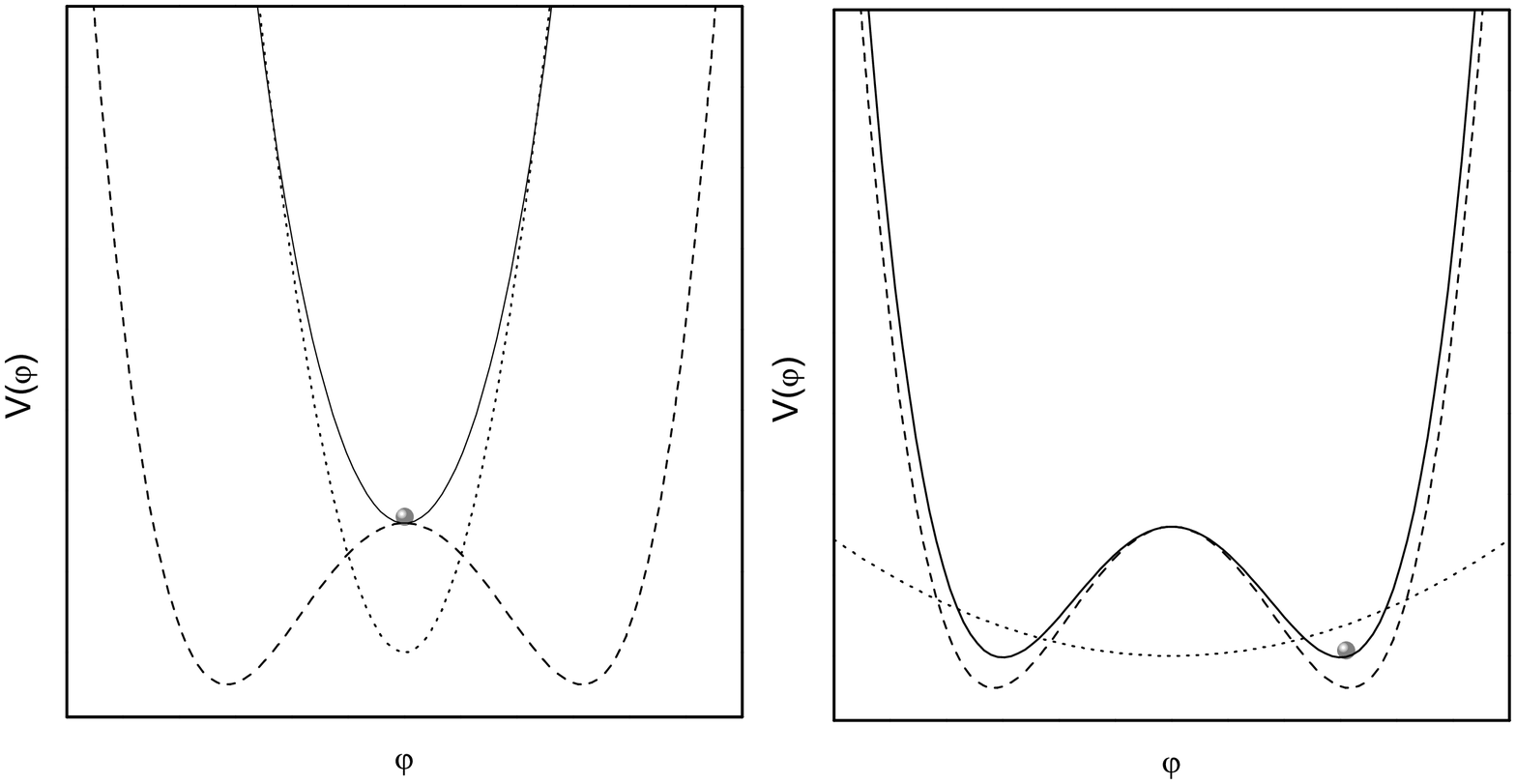}
\caption{An illustration of how the symmetron mechanism works. The dashed, dotted and solid curves are respectively the bare potential $V(\varphi)$ of the symmetron field, the coupling function and the total effective potential $V_{\rm eff}(\varphi)$. {\it Left Panel}: in high matter-density regions the minimum of $V_{\rm eff}(\varphi)$ is where the coupling strength vanishes and so the fifth force is suppressed. {\it Right Panel}: in low matter-density regions the coupling strength does not vanish at the minima of $V_{\rm eff}(\varphi)$, where the symmetron field resides, so a nonzero fifth force takes effect in the structure formation.} \label{fig:symmetron}
\end{figure*}

The initial coupling (taken at $a_{\rm ini}< a_{\rm BBN}$) is the same as in dense matter on Earth and  is related to the cosmological value of $\beta$ today, $\beta(\varphi_0)$, by
\begin{equation}
\vert \beta (\varphi_0)\vert = \vert \beta (\varphi_c)\vert \exp\left[9 A_2\Omega_{m0}H_0^2 \int_{a_{\rm ini}}^1 \frac{\rd a}{a^4 m^2 (a)}\right].
\end{equation}
It is possible to
have a very small coupling in dense matter $\left(\vert \beta (\varphi_c)\vert \ll 1\right)$ for any value of the coupling on cosmological scales $\left(\vert \beta (\varphi_0)\vert\right)$ provided that $A_2 >0$ and that the time variation of $m(a)$ is slow and does not compensate the $1/a^4$ divergence in the integrand. In this situation, the coupling function $\beta$ converges exponentially towards zero: this is the Damour-Polyakov mechanism \cite{dp1994}. The fact that $A_2>0$ guarantees that the minimum of the coupling function $A(\varphi)$ is stable and becomes the minimum of the effective potential which attracts the scalar field \HANS{at late times}. If $A_2<0$, the effect of the coupling is destabilising and implies that $\varphi$ diverges exponentially fast away from $\varphi_ \ast$. 

Alternatively, a  smooth variation of the coupling function to matter in the cosmological background and therefore interesting consequences for the large-scale structure can be achieved when the evolution of the mass of the scalar field compensates  the $1/a^4$ factor in the radiation era and evolves in the matter era. This is obtained for models with
\be
m^2 (a) = 3 A_2 H^2 (a)\mpl^2.
\ee
Indeed, $H(a) \sim a^{-2}$ in the radiation era, which implies that the time variation of $\beta(\varphi)$ between BBN and matter-radiation equality is
\be
\beta(\varphi)=\beta(\varphi_c)\exp\left[3\frac{\Omega_{m0}}{\Omega_{r0}}(a- a_{\rm ini})\right],
\ee
in which $\Omega_r$ is the fractional density for radiation, and in the matter-dominated era
\be
\beta(\varphi)= \beta\left(\varphi_{\rm eq}\right)\left(\frac{a}{a_{\rm eq}}\right)^{{3}},
\ee
in which a subscript $_{\rm eq}$ denotes the value of a quantity at the matter-radiation equality. This is the behaviour of the dilaton models already analysed in \cite{bbdls2011}.

\subsubsection{Symmetron}

The symmetron model was originally described in \cite{hk2010,pospelov}, for which the coupling function and bare potential of the scalar field take the following forms respectively:
\begin{eqnarray}
\label{eq:sym_b}A(\varphi) &=& 1 + \frac{1}{2}\left(\frac{\varphi}{M}\right)^2,\\
\label{eq:sym_v}V(\varphi) &=& V_0-\frac{1}{2}\mu^2\varphi^2+\frac{1}{4}\lambda\varphi^4.
\end{eqnarray}
Here $M\lesssim 10^{-3}M_{\rm Pl}$ is a mass scale and $\mu\sim H_0,\lambda\ll1$ are model parameters. The screening mechanism of the symmetron model is shown in Fig.~\ref{fig:symmetron}. When the matter density is high $\varphi_{\rm min}$ coincides with the minimum of $A(\varphi)$ such that $\beta(\varphi_{\rm min})=0$ and the fifth force vanishes, whilst when matter density is low $\beta(\varphi_{\rm min})\neq0$, resulting in a cosmologically interesting fifth force.

A fundamental property of the symmetron models is that the coupling to matter vanishes identically in dense regions or at redshifts $z>z_ \ast$, and an order-unity coupling is obtained after a transition at a redshift $z_ \ast$ and in the low matter-density regions. In the original symmetron model, this is given by
\be\label{symc}
\beta(a)=  \beta_\star \sqrt{1-\left(\frac{a_ \ast}{a}\right)^3},
\ee
for $z<z_ \ast$ and $\beta=0$ for $z>z_\ast$. Similarly,
\begin{equation}
m(a)=m_\star\sqrt{1-\left(\frac{a_ \ast}{a}\right)^3}.
\ee
\BLED{Notice that for symmetron models a subscript $_{\star}$ denotes the value at far future $\left(a\rightarrow\infty\right)$, and a subscript $_\ast$ means the value at the symmetry breaking, i.e., when $\beta(a)$ becomes nonzero in the cosmological background.}

Using the reconstruction mapping, it is straightforward to find that
\begin{equation}
\varphi(a)=\varphi_\star\sqrt{1-\left(\frac{a_\ast}{a}\right)^3},
\end{equation}
for $z<z_ \ast$ and $\varphi=0$ before. Here we have defined
\be
\varphi_\star\equiv\frac{2\beta_\star\rho_\ast}{m_\star^2\mpl},
\ee
and
\be\label{eq:temp1}
m_\star\equiv\sqrt {2} \mu,\ \
\rho_\ast\equiv\rho_{m0}a_\ast^{-3}.
\ee
The potential for $z<z_\ast$ as a function of $a$ can then be reconstructed, using the technique introduced above, as
\be
V(a)=V_0+\frac{\beta_\star^2\rho_ \ast^2}{2m_\star^2\mpl^2}\left[\left(\frac{a_\ast}{a}\right)^6-1\right].
\ee
The potential as a function of $\varphi$ can then be found to take the form of Eq.~(\ref{eq:sym_v}), with $\mu$ given in Eq.~(\ref{eq:temp1}) and
\be\label{eq:temp2}
\lambda = \frac{\mu^2}{\varphi_\star^2}.
\ee
Meanwhile, $\beta$ as a function of $\varphi$ is reconstructed as
\be
\beta(\varphi)= \frac{\beta_\star}{\varphi_\star}\varphi.
\end{equation}
\BLED{It could be checked that this agrees with Eq.~(\ref{eq:sym_b}), by taking $\beta=\rd\ln A/\rd\varphi\approx\rd A/\rd\varphi$, where the $\approx$ symbol comes from the fact that $A\approx1$.}

\section{Generalised Symmetron and Dilaton Models}

\label{sect:models}

In this section we discuss the  generalisations of the dilaton and symmetron models, and  the effects  of varying the model parameters.

\subsection{Generalised symmetron model}

\label{subsect:symmetron_model}

\subsubsection{Model parameterisation}

The original symmetron model discussed in the previous section only includes one specific potential. As a straightforward generalisation of this idea, let us consider the following $m(a)$ and $\beta(a)$:
\begin{eqnarray}\label{eq:m_new_symmetron}
m(a) &=& \BLED{m_\star}\left[1-\left(\frac{a_\ast}{a}\right)^3\right]^{\BLED{\hat{m}}},\\
\label{eq:beta_new_symmetron}\beta(a) &=& \BLED{\beta_\star}\left[1-\left(\frac{a_\ast}{a}\right)^3\right]^{\BLED{\hat{n}}},
\end{eqnarray}
where \BLED{$\hat{m},\hat{n}$} are two new parameters and not necessarily equal to each other, and \BLED{$(m_\star,\beta_\star)$} are the mass and coupling in vacuum as above. As in \cite{bdl2011}, if the scalar field always follows\footnote{See \cite{bbdlss} for a more detailed discussion on the time-evolution of $\varphi$.} $\varphi_{\rm min}$,  one can obtain the following solution for $\varphi(a)$:
\be\label{eq:varphi_of_a}
\varphi(a)
=\BLED{\varphi_\star}\left[1-\left(\frac{a_\ast}{a}\right)^3\right]^{\BLED{\hat{n}-2\hat{m}+1}},
\ee
where we have defined $\BLED{\varphi_\star} \equiv \frac{3}{\BLED{\hat{n}-2\hat{m}+1}}\Omega_{m}\BLED{\beta_\star}\xi^2a_\ast^{-3}$ \BLED{and from here we will neglect the subscript $_0$ in $\Omega_{m0}$}. Note that Eq.~(\ref{eq:varphi_of_a}) is only valid if $\BLED{\hat{n}-2\hat{m}+1}\neq0$; the case of $\BLED{\hat{n}-2\hat{m}}=-1$ corresponds to a potential that is not bounded below and is therefore not a viable physical model. Again, Eq.~(\ref{eq:varphi_of_a}) is for $a\geq a_\ast$ and for $a<a_\ast$ we have $\varphi(a)=0$.

To study the nonlinear evolution of $\varphi$, we need to know $V_\varphi(\varphi)$ as it appears in the $N$-body equations Eq. (\ref{eq:sf}). Noting  that $\varphi$ increases monotonically with $a$, we find
\begin{eqnarray}
V_\varphi &=& \frac{\d[V(a)]}{\d a}\frac{\d a}{\d\varphi}\nonumber\\
&=& \label{eq:V_varphi_of_a_symmetron}-\BLED{(\hat{n}-2\hat{m}+1)m_\star^2\varphi_\star}\left[1-\left(\frac{a_\ast}{a}\right)^3\right]^{\BLED{\hat{n}}}\nonumber\\
&=& \label{eq:V_varphi_pf_varphi} -\BLED{(\hat{n}-2\hat{m}+1)m_\star^2\varphi_\star}\left(\frac{\varphi}{\BLED{\varphi_\star}}\right)^{\BLED{\frac{\hat{n}}{\hat{n}-2\hat{m}+1}}}\nonumber\\
&& \times \left[1-\left(\frac{\varphi}{\BLED{\varphi_\star}}\right)^{\frac{1}{\BLED{\hat{n}-2\hat{m}+1}}}\right].
\end{eqnarray}
Defining the parameters
\begin{eqnarray}\label{eq:newparam}
M \equiv \BLED{\frac{2\hat{n}-2\hat{m}+2}{\hat{n}-2\hat{m}+1}},\ \ \ \ N\equiv\BLED{\frac{2\hat{n}-2\hat{m}+1}{\hat{n}-2\hat{m}+1}},
\end{eqnarray}
we find that the potential can be written quite simply as
\begin{eqnarray}\label{eq:fullpot}
V(\varphi) = \frac{H_0^2\BLED{\varphi_\star^2}}{\xi^2(M-N)}\left[-\frac{1}{N}\left(\frac{\varphi}{\BLED{\varphi_\star}}\right)^{N} + \frac{1}{M}\left(\frac{\varphi}{\BLED{\varphi_\star}}\right)^{M}\right]. \ \ \ \ \
\end{eqnarray}
In a similar manner, for $a\geq a_\ast$ we get
\begin{eqnarray}\label{eq:beta_of_varphi}
\beta(\varphi) &=& \beta(a(\varphi)) = \BLED{\beta_\star}\left(\frac{\varphi}{\BLED{\varphi_\star}}\right)^{N-1}.
\end{eqnarray}
It is evident that when $N=2$ and $M=4$ we recover the original symmetron model. In what follows we will only consider $M,N$ to be even and positive integers with $M>N$ to avoid having a potential that is unbounded from below.

\subsubsection{Effects of varying model parameters}

\label{subsect:symmetron_effect}

Let us analyse the effects of varying the five model parameters $a_ \ast$, $\BLED{\beta_\star}$, $N,M$ and $\xi$ on structure formation.

As discussed in \cite{bdlw2012}, the modifications of the structure formation at the linear perturbation level is completely determined by the two temporal functions $m(a)$ and $\beta(a)$, from which we can see that:
\begin{enumerate}
\item The strength of the fifth force vanishes for $a<a_ \ast$ and approaches $2\BLED{\beta_\star^2}$ times that of the Newtonian gravity for $a\gg a_ \ast$. Decreasing $a_ \ast$ increases the time during which the fifth force is active thus enhances the matter clustering today.
\item Increasing $\BLED{\beta_\star}$ makes $\beta$ larger at all times, which makes the fifth force stronger and leads to more clustering.
\item \BLED{According to Eq.~(\ref{eq:beta_of_varphi}), increasing $N$ makes $\beta$ smaller because $|\varphi|<|\varphi_\star|$ in general. This can weaken the effect of the fifth force. It is because of this reason that the symmetron screening is more efficient than the chameleon screening with a constant $\beta$ \citep{bdlw2012}.}
\item \BLED{By increasing $M$ the scalar field will make the transition from $\varphi=0$ to $\varphi=\varphi_\star$ much quicker, because then $\varphi^M$ is smaller for small $\varphi$ and so (1) the symmetry in $V_{\rm eff}(\varphi)$ is easier to be broken and (2) $V_{\rm eff}(\varphi)$ becomes steeper from $\varphi=0$ to $\varphi=\varphi_{\star}$.
This leads to a stronger (and earlier kick-in of the) fifth force and thus matter becomes more clustered.}
\item An increase in $\xi$ is equivalent to an increase in the range $\BLED{\lambda_\star}$ of the fifth force since $\BLED{\lambda_\star} \equiv 2998\xi$ Mpc$/h$ \BLED{in vacuum}. This extends the modifications of gravity to larger cosmological scales \BLED{and decreases the exponential factor $e^{-mr}$ of suppression of the fifth force}.
\end{enumerate}


These properties will be investigated in depth using $N$-body simulations below.

\subsection{Generalised dilaton model}

\label{subsect:dilaton_model}

\subsubsection{Model parameterisation}

The environment-dependent dilaton model has already been presented in the previous section.
For the model in \cite{bbds2010} it can be shown that
\begin{eqnarray}\label{eq:m_original_dilaton}
m(a) &=& m_0a^{-\frac{3}{2}},\\
\label{eq:beta_original_dilaton}\beta(a) &=& \beta_0a^{9\Omega_mA_2\xi^2},
\end{eqnarray}
where both $m(a)$ and $\beta(a)$ are power law functions of $a$. If
\begin{eqnarray}\label{eq:m_new_dilaton}
m(a) &=& m_0a^{-r},
\end{eqnarray}
with $r\neq3/2$, then $\beta$ is no longer a power law function of $a$, as we
will see below.

As a straightforward generalisation of the dilaton idea, let us consider a quadratic coupling function $A(\varphi)$ which has a minimum at $\varphi_\ast$. Near $\varphi_\ast$ we have $\beta(\varphi)\approx A_2(\varphi-\varphi_\ast)/M_{\rm Pl}$. Assuming that the dilaton field always follows the minimum of $V_{\rm eff}(\varphi)$, $\varphi_{\rm min}$, one can solve for $\beta(a)$ from an integral \cite{bdl2011}:
\begin{eqnarray}\label{eq:beta_new_dilaton}
\beta(a\leq1) &=& \beta_0\exp\left[9\Omega_mA_2\xi^2\int^a_1a^{2r-4}da\right]\nonumber\\
&=& \beta_0\exp\left[\frac{s}{2r-3}(a^{2r-3}-1)\right],
\end{eqnarray}
in which we have used Eq.~(\ref{eq:m_new_dilaton}) and defined $s\equiv9\Omega_mA_2\xi^2$. Eq.~(\ref{eq:beta_new_dilaton}) is only valid when $r\neq3/2$, while the case of $r=3/2$ corresponds to $m(a)$ and $\beta(a)$ both being non-power-law, which will be studied elsewhere.

As in the symmetron case, we need to have the expression of $V_\varphi(\varphi)$ to study the nonlinear evolution of $\varphi$. For this we will use the relations
\begin{eqnarray}
&&\frac{{\rm d}(\kappa V)}{{\rm d}a}\nonumber\\
&=& -3\frac{\beta^2(a)}{am^2(a)}\frac{\rho^2_m(a)}{M_{\rm Pl}^{4}}\nonumber\\
&=& -27\Omega_m^2\beta_0^2\xi^2H_0^2a^{2r-7}\exp\left[\frac{2s}{2r-3}(a^{2r-3}-1)\right],
\end{eqnarray}
where we have used the expressions of $m(a)$ and $\beta(a)$ given in Eqs.~(\ref{eq:m_new_dilaton},\ref{eq:beta_new_dilaton}), and
\begin{eqnarray}
&&\frac{\d(\sqrt{\kappa}\varphi)}{\d a}\nonumber\\
&=& 3\frac{\beta(a)}{am^2(a)}\frac{\rho_m(a)}{M_{\rm Pl}^{2}}\nonumber\\
&=& 9\Omega_m\beta_0^2\xi^2a^{2r-4}\exp\left[\frac{s}{2r-3}(a^{2r-3}-1)\right].
\end{eqnarray}
Using the above two equations, it is straightforward to find
\begin{eqnarray}
\sqrt{\kappa}V_\varphi &=& \frac{\d[\kappa V(a)]/\d a}{\d(\sqrt{\kappa}\varphi)/\d a}\nonumber\\
&=& \label{eq:V_varphi_of_a_dilaton}-3\Omega_m\beta_0H_0^2\exp\left[\frac{s}{2r-3}(a^{2r-3}-1)\right]a^{-3}\\
&=& \label{eq:V_varphi_of_varphi_dilaton}-3\Omega_mH_0^2\frac{A_2(\varphi-\varphi_\ast)}{M_{\rm Pl}}\nonumber\\
&& \times\left[1+\frac{2r-3}{s}\log\frac{A_2(\varphi-\varphi_\ast)}{M_{\rm Pl}\beta_0}\right]^{-\frac{3}{2r-3}},
\end{eqnarray}
where Eq.~(\ref{eq:V_varphi_of_a_dilaton}) can be used directly when one needs the background value of $V_\varphi(\varphi)$ and Eq.~(\ref{eq:V_varphi_of_varphi_dilaton}) can be used in full nonlinear calculations such as the $N$-body simulations. As in general $A_2(\varphi-\varphi_\ast)/M_{\rm Pl}<\beta_0$, the logarithmic here is negative, and to make sure the last line of Eq.~(\ref{eq:V_varphi_of_varphi_dilaton}) is well defined for any $r$ we should require $r<3/2$. Otherwise the terms in the brackets can be negative when $\varphi\rightarrow\varphi_\ast$, making the power function ill-defined. Because $\varphi$ appears in both $\beta(\varphi)$ and $V_\varphi(\varphi)$ through $\varphi-\varphi_\ast$, without loss of generality, in what follows we take $\varphi_\ast=0$ by a redefinition of $\varphi$.

\subsubsection{Effects of varying model parameters}

\label{subsect:dilaton_effect}

As in the symmetron model, let us first analyse how varying the four parameters $A_2, \beta_0, r$ and $\xi$ affects the structure formation.

\begin{enumerate}
\item Increasing $A_2$ enhances $s=9\Omega_mA_2\xi^2$ and so makes $\beta(a)$ smaller at $a<1$. As $\beta(a)$ controls the strength of the fifth force, this weakens its effect.
\item Increase in $\beta_0$ makes $\beta(a)$ larger at all times, which strengthens the fifth force.
\item The effects of $r$ are two-fold. On the one hand, increasing $r$ makes $m(a)$ larger and therefore the fifth force shorter ranged for $a<1$; on the other hand, it makes $\beta(a)$ larger for $a<1$, provided that $2r-3$ is not very close to $0$, and this strengthens the fifth force. As a result, we expect that this will decrease the matter clustering on large scales but increase it on small scales.
\item An increase in $\xi$ is equivalent to a decrease in $m_0$ and an increase in $s$, which means that both $m(a)$ and $\beta$ become smaller for $a<1$. This increases the matter clustering on large scales and decreases it on small scales. Because of the exponential function in $\beta(a)$, the effect of changing $\xi$ is most significant at early times.
\item There are degeneracies between the different effects. For example, increasing $r$ and decreasing $\xi$ are expected to leave similar imprints on the large-scale structure, as we see below.
\end{enumerate}

Note that the dependence on $\xi$ is quite different from that in the chameleon models with constant coupling $\beta$ \citep{lz2009,lz2010,zlk2011}, and the symmetron model \cite{bdlw2012}. In those cases, increasing $\xi$ decreases  $m(a)$ and therefore increases the range of the fifth force, resulting in more matter clustering.

The above analyses only apply to linear perturbations, the dependence of the fifth force on the dilaton parameters is more complex in the nonlinear regime, and this is best seen from the two functions $\beta(\varphi)$ and $V_\varphi(\varphi)$, which govern the nonlinear equations (see above):
\begin{enumerate}
\item Increasing $A_2$ implies that the parabolic function $A(\varphi)$ becomes steeper near its minimum at $\varphi=\varphi_\ast$, and this makes it harder for the scalar field to roll away from $\varphi_\ast$, where $\beta(\varphi)=0$. This weakens the fifth force.
\item Increasing $\beta_0$ makes $A_2(\varphi-\varphi_\ast)/M_{\rm Pl}\beta_0$ closer to zero and therefore $|V_\varphi(\varphi)|$ larger. This means that $V(\varphi)$ becomes steeper, making it easier for the scalar field to roll away from $\varphi_\ast$ where $\beta(\varphi)=0$ and therefore strengthening the fifth force.
\item If $2r-3$ is not too close to zero, increasing $r$ towards $3/2$ makes $|V_\varphi(\varphi)|$ larger according to Eq.~(\ref{eq:V_varphi_of_varphi_dilaton}) and so makes it easier for the scalar field to roll away from $\varphi_\ast$ where $\beta(\varphi)=0$. This strengthens the fifth force.
\item Similarly, increasing $\xi$ (therefore $s$) makes $V(\varphi)$ shallower and the fifth force weaker. Meanwhile, the scalar field becomes less massive and therefore less likely to follow the local minimum of $V_{\rm eff}$ which is determined by the matter density field and more likely to take larger values -- this could give rise to a larger value of $\beta$ and therefore a stronger fifth force.
\end{enumerate}

\section{The $N$-body Simulations}

\label{sect:nbody_eqns}

\subsection{Equations in code units}

In this section we derive the equations used in the $N$-body simulations, namely, the Poisson equation for the gravitational potential and the EOM governing the dynamics of the scalar field. For the sake of completeness we first describe the code units used in these equations.
The code units used in our code are based on (but not exactly the same as) the supercomoving coordinates of \cite{ms1998}. They can be summarised as follows (tilded quantities are expressed in code units):
\begin{eqnarray}
\tilde{x}\ =\ \frac{x}{aB},\ \ \ \tilde{\rho}\ =\ \frac{\rho a^3}{\rho_c\Omega_m},\ \ \ \tilde{v}\ =\ \frac{av}{BH_0},\nonumber\\
\tilde{\phi}\ =\ \frac{a^2\phi}{(BH_0)^2},\ \ \ d\tilde{t}\ =\ H_0\frac{dt}{a^2},\ \ \ \tilde{c}\ =\ \frac{c}{BH_0}.\nonumber
\end{eqnarray}
In the above $x$ is the comoving coordinate, $\rho_c$ is the critical density today, $\Omega_m$ the fractional energy density for matter today, $v$ the particle velocity, $\phi$ the gravitational potential and $c$ the speed of light. In addition, $B$ is the size of the simulation box in unit of $h^{-1}$Mpc and $H_0$ the Hubble expansion rate today in units of $100h$~km/s/Mpc. Note that with these conventions the average matter density is $\BLED{\tilde{\bar{\rho}}}=1$ at all times. All the newly defined quantities are dimensionless.

Using the code units defined above, the Poisson equation Eq.~(\ref{eq:newton}) becomes
\begin{eqnarray}\label{eq:newton_nbody}
\tilde{\nabla}^2\tilde{\phi} &\approx& \frac{3}{2}\Omega_ma\left(\tilde{\rho}-1\right).
\end{eqnarray}
Note that the Poisson equations for both the symmetron and the dilaton cases are unchanged compared to the case of standard GR, because we have neglected the contribution from the scalar field to the source term. In what follows, we introduce the symmetron and dilaton versions of the scalar field equation, i.e., Eq~(\ref{eq:sf}).

\subsubsection{The symmetron case}

Throughout the cosmic history, the symmetron field has a small magnitude, i.e., $|\varphi|/M_{\rm Pl}\ll1$. To guarantee the numerical accuracy, instead of solving $\varphi$ itself, we solve for a newly-defined variable $\tilde{\varphi}\equiv\varphi/\BLED{\varphi_\star}$. This variable is constrained by $0\leq |\tilde{\varphi}| \leq 1$ everywhere.
The symmetron equation of motion Eq.~(\ref{eq:sf}) becomes
\begin{eqnarray}\label{eq:sf_nbody_symmetron}
\tilde{\nabla}^2\tilde{\varphi} &\approx& \frac{a^2}{(M-N)\tilde{c}^2\xi^2}\tilde{\varphi}^{N-1}\left[\tilde{\rho}\left(\frac{a_\ast}{a}\right)^3-1\right]\nonumber\\
&& +\frac{a^2}{(M-N)\tilde{c}^2\xi^2}\tilde{\varphi}^{M-1}.
\end{eqnarray}

\subsubsection{The dilaton case}

Similarly, the dilaton field $\varphi$ is generally very small ($\varphi\ll M_{\rm Pl}$) and should be positive
(otherwise the logarithmic in Eq.~(\ref{eq:V_varphi_of_varphi_dilaton}) is ill-defined). This means that the numerical value of $\varphi$ can easily go negative in the relaxation procedure, leading to the failure of convergence. To avoid this problem, we follow \cite{o2008, lz2009} and use a newly-defined variable $u=\log(\varphi/M_{\rm Pl})$ instead of $\varphi$ itself. During the cosmic evolution $|u|$ remains $\mathcal{O}(1)\sim\mathcal{O}(10)$, compared to the several orders-of-magnitude span of $\varphi$, making it easier to handle the numerical errors.

 After some simplification, the dilaton equation of motion Eq.~(\ref{eq:sf}) becomes
\begin{eqnarray}\label{eq:sf_nbody_dilaton}
\tilde{\nabla}^2e^u &\approx& \frac{3}{\tilde{c}^2}\Omega_mA_2\tilde{\rho}e^ua^{-1}\\
&&-\frac{3}{\tilde{c}^2}\Omega_mA_2e^u\left[a^{2r-3}+\frac{2r-3}{s}\log\frac{e^u}{\bar{\varphi}}\right]^{-\frac{3}{2r-3}}a^2.\nonumber
\end{eqnarray}

\subsection{The discretised equations}

\label{sect:discrete_eqns}

Evidently, to put the above equations into the $N$-body code one must discretise them. For the Poisson equation we have
\begin{eqnarray}\label{eq:newton_discrete}
\frac{1}{h^2}\big[\tilde{\phi}_{i+1,j,k}+\tilde{\phi}_{i-1,j,k}+\tilde{\phi}_{i,j+1,k}+\tilde{\phi}_{i,j-1,k}+\tilde{\phi}_{i,j,k+1}\nonumber\\
+\tilde{\phi}_{i,j,k-1}-6\tilde{\phi}_{i,j,k}\big] = \frac{3}{2}\Omega_ma\left(\tilde{\rho}_{i,j,k}-1\right),\ \ \
\end{eqnarray}
where $\tilde{\phi}_{i,j,k}$ is the value of $\tilde{\phi}$ in the grid cell with index $(i,j,k)$.

\subsubsection{Symmetron equation of motion}

The discrete version of the nonlinear symmetron EOM can be obtained similarly:
\begin{eqnarray}\label{eq:sf_discrete_symmetron}
L^h(\tilde{\varphi}_{i,j,k}) &=& 0,
\end{eqnarray}
where the operator $L^h(\tilde{\varphi}_{i,j,k})$ is defined as
\begin{eqnarray}\label{eq:op_symmetron}
L^h(\tilde{\varphi}_{i,j,k}) &\equiv& \frac{1}{h^2}\big[\tilde{\varphi}_{i+1,j,k}+\tilde{\varphi}_{i-1,j,k}+\tilde{\varphi}_{i,j+1,k}+\tilde{\varphi}_{i,j-1,k}\nonumber\\
&&+\tilde{\varphi}_{i,j,k+1}+\tilde{\varphi}_{i,j,k-1}-6\tilde{\varphi}_{i,j,k}\big]\nonumber\\
&&-\frac{a^2}{(M-N)\tilde{c}^2\xi^2}\tilde{\varphi}_{i,j,k}^{N-1}\left[\tilde{\rho}_{i,j,k}\frac{a^3_\ast}{a^3}-1\right]\nonumber\\
&&-\frac{a^2}{(M-N)\tilde{c}^2\xi^2}\tilde{\varphi}_{i,j,k}^{M-1}.
\end{eqnarray}
Eq.~(\ref{eq:sf_discrete_symmetron}) is solved using the nonlinear Gauss-Seidel relaxation, which can be summarised as
\begin{eqnarray}\label{eq:ngs_symmetron}
\tilde{\varphi}^{h,\rm new}_{i,j,k} &=& \tilde{\varphi}^{h,\rm old}_{i,j,k}-\frac{L^h\big(\tilde{\varphi}^{h,\rm old}_{i,j,k}\big)}{\frac{\partial L^h\left(\tilde{\varphi}^{h,\rm old}_{i,j,k}\right)}{\partial\tilde{\varphi}^{h,\rm old}_{i,j,k}}},
\end{eqnarray}
where
\begin{eqnarray}\label{eq:dop_symmetron}
\frac{\partial L^h\left(\tilde{\varphi}^{h}_{i,j,k}\right)}{\partial\tilde{\varphi}^{h}_{i,j,k}} &=& -\frac{6}{h^2}-\frac{(N-1)a^2}{(M-N)\tilde{c}^2\xi^2}\tilde{\varphi}_{i,j,k}^{N-2}\left[\tilde{\rho}_{i,j,k}\frac{a^3_\ast}{a^3}-1\right]\nonumber\\
&&-\frac{(M-1)a^2}{(M-N)\tilde{c}^2\xi^2}\tilde{\varphi}_{i,j,k}^{M-2}.
\end{eqnarray}

In practice, Eqs.~(\ref{eq:op_symmetron},\ref{eq:dop_symmetron}) must be modified at the boundaries of refinements for the multigrid implementation, as is the case of the Poisson equation. Ref.~\cite{ecosmog} gives a detailed review of all the technical details involved in the $N$-body code implementation: interested readers are referred to that paper.

\subsubsection{Dilaton equation of motion}

The discrete version of the nonlinear dilaton equation can be obtained similarly:
\begin{eqnarray}\label{eq:sf_discrete_dilaton}
L^h(u_{i,j,k}) &=& 0,
\end{eqnarray}
where the operator $L^h(u_{i,j,k})$ defined as
\begin{widetext}
\begin{eqnarray}\label{eq:op_dilaton}
L^h(u_{i,j,k}) &\equiv& \frac{1}{h^2}\left[b_{i+\frac{1}{2},j,k}u_{i+1,j,k}-u_{i,j,k}\left(b_{i+\frac{1}{2},j,k}+b_{i-\frac{1}{2},j,k}\right)+b_{i-\frac{1}{2},j,k}u_{i-1,j,k}\right]\nonumber\\
&&+\frac{1}{h^2}\left[b_{i,j+\frac{1}{2},k}u_{i,j+1,k}-u_{i,j,k}\left(b_{i,j+\frac{1}{2},k}+b_{i,j-\frac{1}{2},k}\right)+b_{i,j-\frac{1}{2},k}u_{i,j-1,k}\right]\nonumber\\
&&+\frac{1}{h^2}\left[b_{i,j,k+\frac{1}{2}}u_{i,j,k+1}-u_{i,j,k}\left(b_{i,j,k+\frac{1}{2}}+b_{i,j,k-\frac{1}{2}}\right)+b_{i,j,k-\frac{1}{2}}u_{i,j,k-1}\right]\nonumber\\
&&+\frac{3}{\tilde{c}^2}\Omega_mA_2a^2e^{u_{i,j,k}}\left[a^{2r-3}+\frac{2r-3}{s}\frac{u_{i,j,k}}{\bar{\varphi}}\right]^{-\frac{3}{2r-3}} - \frac{3}{\tilde{c}^2}\Omega_mA_2\tilde{\rho}_{i,j,k}a^{-1}e^{u_{i,j,k}}.
\end{eqnarray}
\end{widetext}
Here $b\equiv\partial e^u/\partial u = e^u$,
\begin{eqnarray}
b_{i+\frac{1}{2},j,k} &\equiv& \frac{1}{2}\left(b_{i+1,j,k}+b_{i,j,k}\right),\nonumber\\
b_{i-\frac{1}{2},j,k} &\equiv& \frac{1}{2}\left(b_{i,j,k}+b_{i-1,j,k}\right),~\cdots\nonumber
\end{eqnarray}
and $h$ is the length of the cell in the numerical simulation mesh.

Eq.~(\ref{eq:sf_discrete_dilaton}) is solved using the nonlinear Gauss-Seidel relaxation as well, which can be summarised as
\begin{eqnarray}\label{eq:ngs_dilaton}
u^{h,\rm new}_{i,j,k} &=& u^{h,\rm old}_{i,j,k}-\frac{L^h\big(u^{h,\rm old}_{i,j,k}\big)}{\frac{\partial L^h\left(u^{h,\rm old}_{i,j,k}\right)}{\partial u^{h,\rm old}_{i,j,k}}},
\end{eqnarray}
where
\begin{widetext}
\begin{eqnarray}\label{eq:dop_dilaton}
\frac{\partial L^h\left(u_{i,j,k}\right)}{\partial u_{i,j,k}} &=& \frac{\tilde{c}^2}{2h^2}b_{i,j,k}\big[u_{i+1,j,k}+u_{i-1,j,k}+u_{i,j+1,k}+u_{i-1,j,k}+u_{i,j,k+1}+u_{i,j,k-1}-6u_{i,j,k}\big]\nonumber\\
&&-\frac{\tilde{c}^2}{2h^2}\big[b_{i+1,j,k}+b_{i-1,j,k}+b_{i,j+1,k}+b_{i,j-1,k}+b_{i,j,k+1}+b_{i,j,k-1}+6b_{i,j,k}\big]\nonumber\\
&&+3\Omega_mA_2e^u_{i,j,k}a^2\left[a^{2r-3}+\frac{2r-3}{s}\frac{u_{i,j,k}}{\bar{\varphi}}\right]^{-\frac{3}{2r-3}}-\frac{1}{\xi^2}e^{u_{i,j,k}}a^2\left[a^{2r-3}+\frac{2r-3}{s}\frac{u_{i,j,k}}{\bar{\varphi}}\right]^{-\frac{2r}{2r-3}}\nonumber\\
&&-3\Omega_mA_2\tilde{\rho}a^{-2}e^{u_{i,j,k}}.
\end{eqnarray}
\end{widetext}

Again, Eqs.~(\ref{eq:op_dilaton}) and (\ref{eq:dop_dilaton}) must be modified at the boundaries of refinements for the multigrid implementation, as is the case of the Poisson equation.

\section{Code Tests}

\label{sect:code_tests}

In this section we present the results of code tests we have performed to show that our symmetron and dilaton equation solvers work well. To lighten the notation, throughout this section we use the units $M_{\rm Pl}=1$.

\begin{table}
\caption{The parameter values for the six models used in the symmetron code test.}
\begin{tabular}{@{}lcccc}
\hline\hline
\ model\ \ \ \ \ \ \ \ \ \ \ \ & $a_\ast$\ \ \ \ \ \ \ \ \ \ \ \ & $\beta_0$\ \ \ \ \ \ \ \ \ \ \ \ & $(N,M)$\ \ \ \ \ \ \ \ \ \ \ \ & $\xi$\ \ \ \ \ \ \ \ \ \ \ \ \\
\hline
\ \ \ \ \ a\ \ \ \ \ \ \ \ & $0.5$\ \ \ \ \ \ \ \ \ \ \ \ & $0.5$\ \ \ \ \ \ \ \ \ \ \ \ & $(2,4)$\ \ \ \ \ \ \ \ \ \ \ \ & $0.001$\ \ \ \ \ \ \ \ \ \ \ \ \\
\ \ \ \ \ b\ \ \ \ \ \ \ \ & $0.2$\ \ \ \ \ \ \ \ \ \ \ \ & $0.5$\ \ \ \ \ \ \ \ \ \ \ \ & $(2,4)$\ \ \ \ \ \ \ \ \ \ \ \ & $0.001$\ \ \ \ \ \ \ \ \ \ \ \ \\
\ \ \ \ \ c\ \ \ \ \ \ \ \ & $0.5$\ \ \ \ \ \ \ \ \ \ \ \ & $1.0$\ \ \ \ \ \ \ \ \ \ \ \ & $(2,4)$\ \ \ \ \ \ \ \ \ \ \ \ & $0.001$\ \ \ \ \ \ \ \ \ \ \ \ \\
\ \ \ \ \ d\ \ \ \ \ \ \ \ & $0.5$\ \ \ \ \ \ \ \ \ \ \ \ & $0.5$\ \ \ \ \ \ \ \ \ \ \ \ & $(2,6)$\ \ \ \ \ \ \ \ \ \ \ \ & $0.001$\ \ \ \ \ \ \ \ \ \ \ \ \\
\ \ \ \ \ e\ \ \ \ \ \ \ \ & $0.5$\ \ \ \ \ \ \ \ \ \ \ \ & $0.5$\ \ \ \ \ \ \ \ \ \ \ \ & $(2,4)$\ \ \ \ \ \ \ \ \ \ \ \ & $0.0005$\ \ \ \ \ \ \ \ \ \ \ \ \\
\ \ \ \ \ f \ \ \ \ \ \ \ \ & $0.5$\ \ \ \ \ \ \ \ \ \ \ \  & $0.5$\ \ \ \ \ \ \ \ \ \ \ \  & $(2,4)$\ \ \ \ \ \ \ \ \ \ \ \  & $0.002$\ \ \ \ \ \ \ \ \ \ \ \ \\
\hline\hline
\end{tabular}\label{tab:test_models_symmetron}
\end{table}

\begin{table}
\caption{The parameter values for the five models used in the dilaton code test.}
\begin{tabular}{@{}lcccc}
\hline\hline
\ model\ \ \ \ \ \ \ \ \ \ \ \ & $A_2$\ \ \ \ \ \ \ \ \ \ \ \ & $\beta_0$\ \ \ \ \ \ \ \ \ \ \ \ & $r$\ \ \ \ \ \ \ \ \ \ \ \ & $\xi$\ \ \ \ \ \ \ \ \ \ \ \ \\
\hline
\ \ \ \ \ a\ \ \ \ \ \ \ \ & $5\times10^5$\ \ \ \ \ \ \ \ \ \ \ \ & $0.5$\ \ \ \ \ \ \ \ \ \ \ \ & $1$\ \ \ \ \ \ \ \ \ \ \ \ & $0.001$\ \ \ \ \ \ \ \ \ \ \ \ \\
\ \ \ \ \ b\ \ \ \ \ \ \ \ & $1\times10^6$\ \ \ \ \ \ \ \ \ \ \ \ & $0.5$\ \ \ \ \ \ \ \ \ \ \ \ & $1$\ \ \ \ \ \ \ \ \ \ \ \ & $0.001$\ \ \ \ \ \ \ \ \ \ \ \ \\
\ \ \ \ \ c\ \ \ \ \ \ \ \ & $5\times10^5$\ \ \ \ \ \ \ \ \ \ \ \ & $1.0$\ \ \ \ \ \ \ \ \ \ \ \ & $1$\ \ \ \ \ \ \ \ \ \ \ \ & $0.001$\ \ \ \ \ \ \ \ \ \ \ \ \\
\ \ \ \ \ d\ \ \ \ \ \ \ \ & $5\times10^5$\ \ \ \ \ \ \ \ \ \ \ \ & $0.5$\ \ \ \ \ \ \ \ \ \ \ \ & $0$\ \ \ \ \ \ \ \ \ \ \ \ & $0.001$\ \ \ \ \ \ \ \ \ \ \ \ \\
\ \ \ \ \ e\ \ \ \ \ \ \ \ & $5\times10^5$\ \ \ \ \ \ \ \ \ \ \ \ & $0.5$\ \ \ \ \ \ \ \ \ \ \ \ & $1$\ \ \ \ \ \ \ \ \ \ \ \ & $0.002$\ \ \ \ \ \ \ \ \ \ \ \ \\
\hline\hline
\end{tabular}\label{tab:test_models_dilaton}
\end{table}

There are five parameters for the generalised symmetron model, namely $a_\ast,\beta_0,N,M$ and $\xi$, and we set $N=2$ and test the code for 6 models summarised in table~\ref{tab:test_models_symmetron}. There are 4 parameters for the generalised dilaton model, namely $A_2,\beta_0,r$ and $\xi$ (note that $s$ can be calculated when $A_2$ and $\xi$ are given, and is therefore not an independent model parameter), and we test the code for 5 models summarised in table~\ref{tab:test_models_dilaton}.

\subsection{Homogeneous matter density field}

In a universe with a homogeneous density, the symmetron field $\varphi$ should \HANS{exactly take its} background value $\bar{\varphi}$, namely
\begin{eqnarray}\label{eq:varphi_background_symmetron}
\bar{\varphi}(a) &=& \BLED{\varphi_\star}\left[1-\left(\frac{a_\ast}{a}\right)^3\right]^{\frac{1}{M-N}},
\end{eqnarray}
everywhere. Thus, as the simplest test of the symmetron equation solver, one can show that in such a homogeneous field, given some random initial guess of $\varphi$ on the cells of the simulation mesh, after a reasonable number of Gauss-Seidel relaxation sweeps, the solutions all converge to the above background value. Such simple test have been used previously in \cite{bbdls2011,dlmw2012,ecosmog} to show that the solver for extra degrees of freedom works correctly.

\begin{figure}
\includegraphics[scale=0.36]{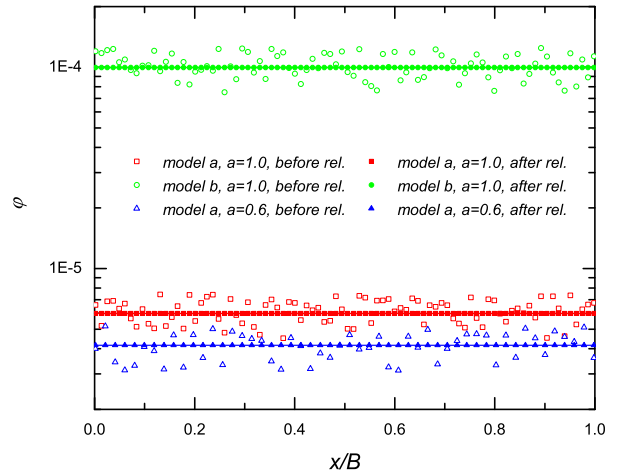}
\caption{(Colour online) Test of the solver for the symmetron equation in a constant matter density field. Only results in the cells along the $x$-axis are shown, and the $x$-coordinate is rescaled by the size of the simulation box so that $x\in[0,1]$. Results for three models as explained in the legend have been shown (the empty symbols), the final answer corresponding to which are filled symbols of the same type and colour. The horizontal lines with the same colours are the exact analytical solution.} \label{fig:test_const_dens_symmetron}
\end{figure}

We have performed this test for all the six symmetron models summarised in Table~\ref{tab:test_models_symmetron}. The result is shown in Fig.~\ref{fig:test_const_dens_symmetron}, where we plot the values of $\varphi/M_{\rm Pl}$ in the cells in the $x$-direction, before and after the Gauss-Seidel relaxation; for clarity we have only shown the results for models a and b at $a=1.0$ and model a at $a=0.6$. We can see that the final solution agrees with the analytical result (the horizontal lines) very well (see figure caption for more details).

We have also tested the code for a model with $a_\ast=0.5$ at $a=0.4$. In this case the symmetry of $V_{\rm eff}(\varphi)$ has not been broken yet, and we expect that $\varphi$ vanishes everywhere. This is confirmed by the tests (which are not shown here).

For the dilaton model, the field $\varphi$ also takes exactly its background value $\bar{\varphi}$, given by
\begin{eqnarray}\label{eq:varphi_background_dilaton}
\bar{\varphi}(a) &=& \frac{\beta_0}{A_2}e^{-\frac{s}{2r-3}}\exp\left[\frac{s}{2r-3}a^{2r-3}\right],
\end{eqnarray}
everywhere in a homogeneous universe.

\begin{figure}
\includegraphics[scale=0.36]{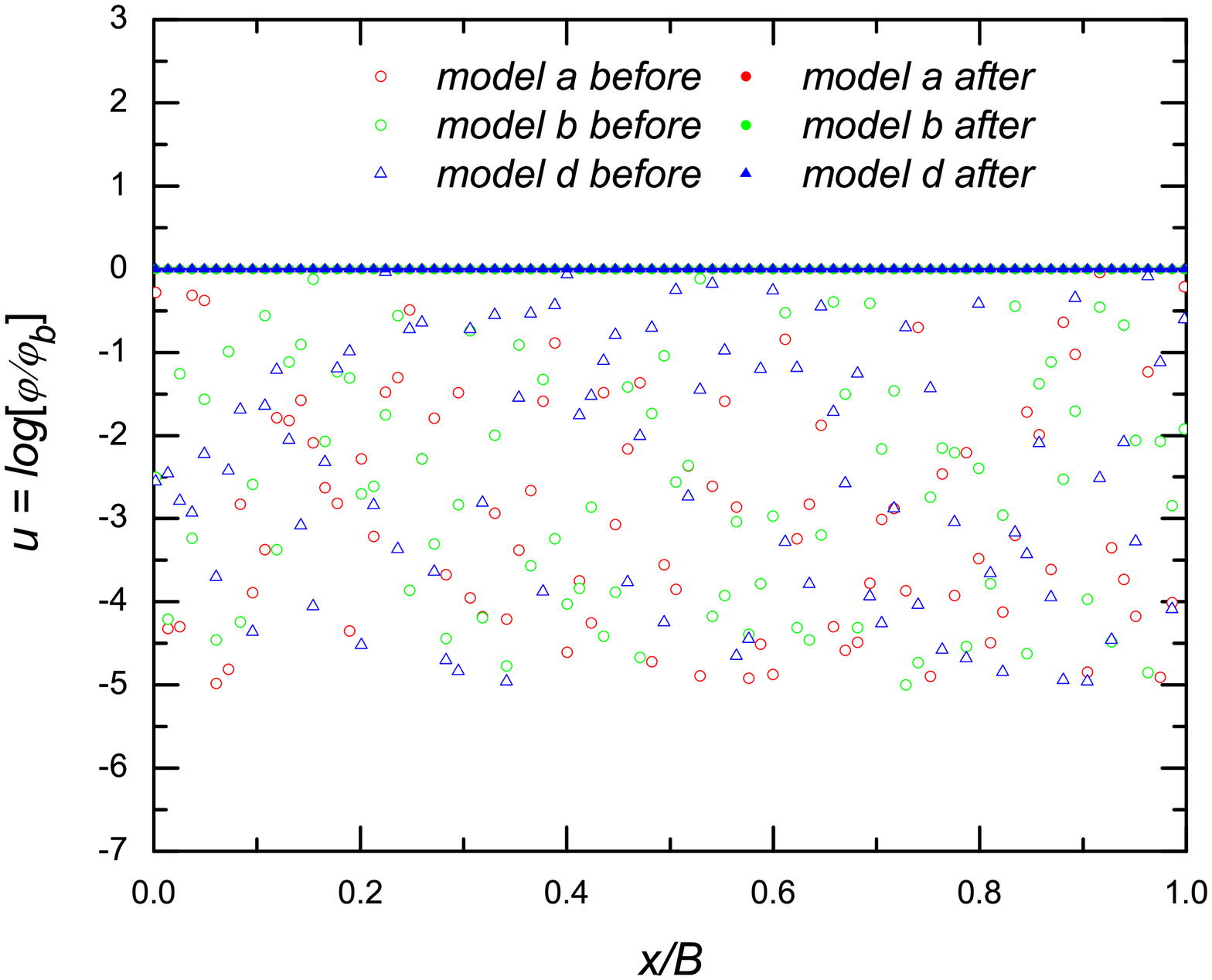}
\caption{(Colour online) \BLED{Similar to Fig.~\ref{fig:test_const_dens_symmetron}, but for the dilaton model. For clarity only the results of models a, b, d (as indicated in the legend) are shown: the initial guesses are represented by the empty symbols and the numerical solutions are denoted by filled symbols of the same type and colour.} Note that, instead of $\log(\varphi)$, we have shown $\log(\varphi/\bar{\varphi})$. The horizontal lines with the same colours are the exact analytical solution, which is zero identically.} \label{fig:test_const_dens_dilaton}
\end{figure}

We have performed this test for three of the five models summarised in Table~\ref{tab:test_models_dilaton}. The results are shown in Fig.~\ref{fig:test_const_dens_dilaton}, where we plot the values of $\log(\varphi/\bar{\varphi})$ in the cells in the $x$-direction, both before and after the relaxation. For clarity we have only shown the results at $a=1.0$. It can be seen that the final solution agrees with the analytical result (the horizontal lines) very well (see figure caption for more details). We have also tested our code at $a\neq1.0$ and found the same good agreement.

\subsection{Point mass}

As a second test of our symmetron equation solver, let us consider the solution of $\varphi$ around a point mass at the origin, for which case we have an analytical solution which is accurate except for the regions very close to the mass. Such a test has been used previously in \cite{o2008,bbdls2011,ecosmog}.

Following \cite{o2008}, we construct the point-mass density field as (hereafter $\delta_{i,j,k}\equiv\tilde{\rho}_{i,j,k}-1$)
\begin{equation}
\label{eq:point_mass}
\delta_{i,j,k} = \left\{%
\begin{array}{ll}
10^{-4}\left(N^3-1\right), & \hbox{$i=j=k=0$;} \\
-10^{-4}, & \hbox{otherwise.} \\
\end{array}%
\right.
\end{equation}
in which $i,j,k$ are respectively the cell indices in the $x,y,z$ direction. In the test we use a cubic box with size $250h^{-1}$Mpc and 256 grid cells in each direction. We have done this test for all six models of table~\ref{tab:test_models_symmetron} at $a=1$.

On the other hand, the analytical solution can be obtained approximately by solving the equation
\begin{eqnarray}\label{eq:linearised_eqn}
\nabla^2\delta\varphi &\approx& m^2\delta\varphi
\end{eqnarray}
in which the effective mass of the scalar field $\delta\varphi=\varphi-\bar{\varphi}$ is $m^2=\xi^2H_0^2$. The analytical solution is
\begin{eqnarray}
\delta\varphi &\propto& \frac{1}{r}\exp(-mr),
\end{eqnarray}
with $r$ the distance from the point mass.

\begin{figure}
\includegraphics[scale=0.36]{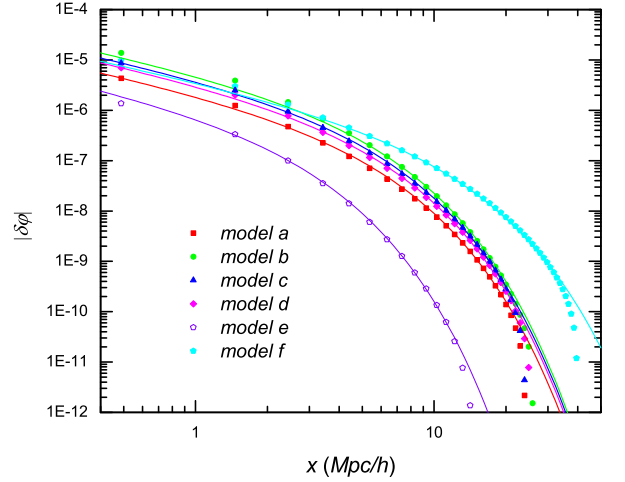}
\caption{(Colour online) The  solution to $\delta\varphi\equiv\varphi-\bar{\varphi}$ around a point mass constructed according to Eq.~(\ref{eq:point_mass}), for the six test symmetron models in Table~\ref{tab:test_models_symmetron} (see the legend). The solid curves with the same colours are the corresponding analytical approximations which are accurate far from the point mass. Only solutions along the $x$-axis are shown.} \label{fig:test_point_mass_symmetron}
\end{figure}

Fig.~\ref{fig:test_point_mass_symmetron} shows the comparison between the numerical solutions to $\delta\varphi$ along the $x$-axis (symbols) and analytical solutions (solid curves) for the symmetron models, and we can see that the two agree very well in all cases. The discrepancies at small $x$ is because the linearisation procedure in deriving Eq.~(\ref{eq:linearised_eqn}) is not accurate and the discrepancy at big $x$ is because the size of $\delta\varphi$ has reached  the level of the discretisation error \cite{o2008}. Fig.~\ref{fig:test_point_mass_dilaton} shows the comparison for the dilaton models, and once again we find excellent agreements.

\begin{figure}
\includegraphics[scale=0.36]{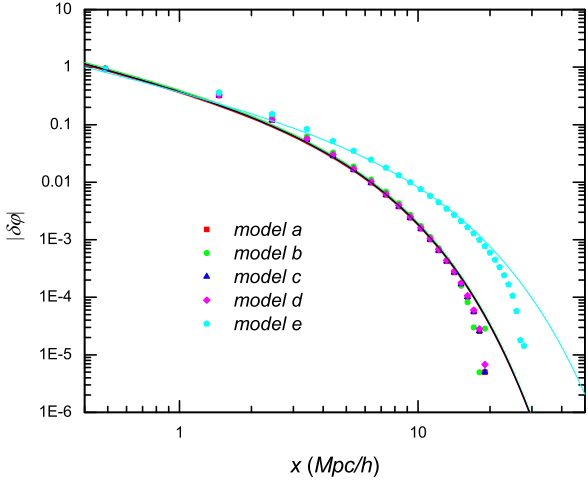}
\caption{(Colour online) The  solution to $\delta\varphi\equiv\varphi-\bar{\varphi}$ around a point mass constructed according to Eq.~(\ref{eq:point_mass}), for the five test dilaton models in Table~\ref{tab:test_models_dilaton} (see the legend). The solid curves with the same colours are the corresponding analytical approximations which are accurate far from the point mass. Only solutions along the $x$-axis are shown.} \label{fig:test_point_mass_dilaton}
\end{figure}

\subsection{Sine density field}

As our third test, let us consider the sine density field introduced in \cite{o2008}, which (after some modification to account for the code units) in the symmetron case is given by
\begin{eqnarray}\label{eq:sine_dens_symmetron}
\left(\frac{a_ \ast}{a}\right)^3\tilde{\rho}(x) &=& 1+\left[\frac{2\pi\tilde{c}\xi}{a}\right]^2\frac{(M-N)\sin(2\pi x)}{[2-\sin(2\pi x)]^{N-1}}\nonumber\\
&-& [2-\sin(2\pi x)]^{M-N},
\end{eqnarray}
where $x$ is rescaled so that $x\in[0,1]$. We consider only the $x$-dependence, which is equivalent to a one-dimensional configuration. The solution to this density field can be analytically worked out to be\footnote{More exactly speaking, we specify the solution we want the code to reproduce and then use the EOM to calculate the corresponding density field that gives rise to this solution.},
\begin{eqnarray}
\varphi(x) &=& \BLED{\varphi_\star}[2-\sin(2\pi x)].
\end{eqnarray}

\begin{figure}
\includegraphics[scale=0.36]{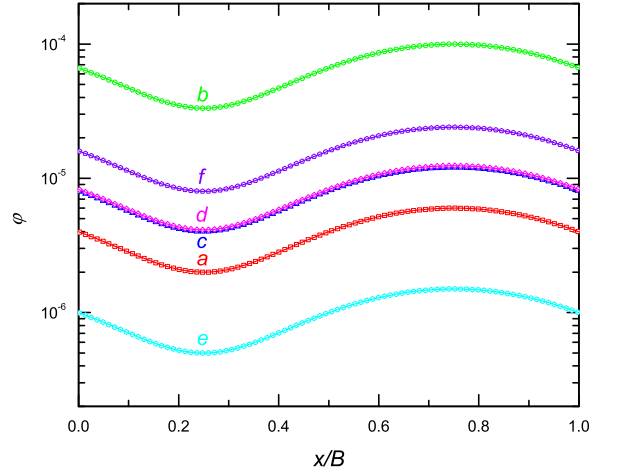}
\caption{(Colour online) Solutions of $\varphi$ in a one-dimensional ($x$-direction) sine density field constructed using Eq.~(\ref{eq:sine_dens_symmetron}), for the six test symmetron models (as indicated besides the curves). The solid curves with same colour are the corresponding analytical results and the symbols are the numerical solutions. A simulation box with side length of $250h^{-1}$Mpc and 256 grid cells on each side is used in the computation. $x$ is rescaled so that $x/B\in[0,1]$.} \label{fig:test_sine_dens_symmetron}
\end{figure}

Fig.~\ref{fig:test_sine_dens_symmetron} shows the symmetron test results for the sine density field given above, at $a=1$ and for the six models listed in Table~\ref{tab:test_models_symmetron}. It can be seen that the numerical solutions (symbols) agree with the analytical solutions (solid curves) very well.

Similarly, for the dilaton field let us consider the following density field
\begin{eqnarray}\label{eq:sine_dens_dilaton}
\tilde{\rho}(x) &=& \frac{\tilde{c}^2a}{\Omega_mA_2}\frac{(2\pi)^2}{3}\frac{\sin(2\pi x)}{2-\sin(2\pi x)}\\
&&+\left[a^{2r-3}+\frac{2r-3}{s}\log\left[\frac{2-\sin(2\pi x)}{3}\right]\right]^{-\frac{3}{2r-3}}a^3,\nonumber
\end{eqnarray}
in which $x$ is rescaled such that $x\in[0,1]$. The solution to this density field can be analytically worked out to be,
\begin{eqnarray}
\varphi(x) &=& \frac{1}{3}\bar{\varphi}\left[2-\sin(2\pi x)\right].
\end{eqnarray}

\begin{figure}
\includegraphics[scale=0.36]{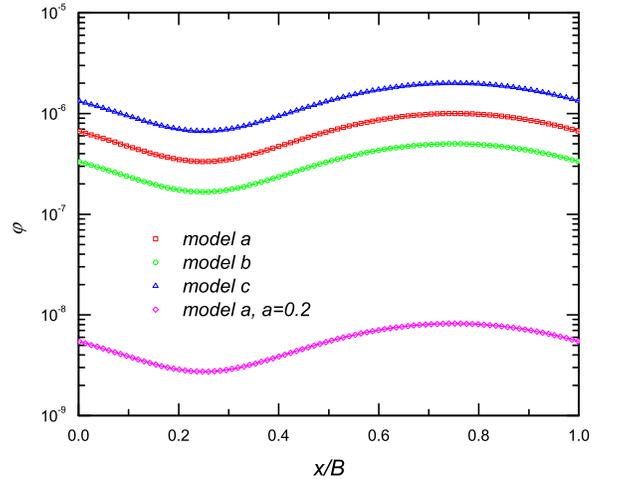}
\caption{(Colour online) Solutions of $\varphi$ in a one-dimensional ($x$-direction) sine density field constructed using Eq.~(\ref{eq:sine_dens_dilaton}), for three test dilaton models (a, b, c) at $a=1.0$ and model a at $a=0.2$ (as indicated besides the curves). The solid curves are the corresponding analytical results and the symbols are the numerical solutions. A simulation box with side length of $250h^{-1}$Mpc and 256 grid cells on each side is used in the computation. $x$ is rescaled so that $x/B\in[0,1]$.} \label{fig:test_sine_dens_dilaton}
\end{figure}

Fig.~\ref{fig:test_sine_dens_dilaton} shows the dilaton test results for the sine density field given above, at $a=1.0$ for models a, b, c and at $a=0.2$ for model a listed in Table~\ref{tab:test_models_dilaton}. As in the symmetron case, the agreement is very good.

\subsection{Gaussian density field}

The last test on the regular (i.e., unrefined) grid uses a Gaussian type density configuration. Again, here we only consider one dimension, and for the symmetron case the density field is specified as
\begin{eqnarray}\label{eq:gaussian_dens_symmetron}
\left(\frac{a_ \ast}{a}\right)^3\tilde{\rho}(x) &=& 1 + \left(\frac{\tilde{c}\xi}{a}\right)^2\frac{\alpha(M-N)(x-0.5)^2/W^2}{\left(1-\alpha\exp\left[-\frac{(x-0.5)^2}{W^2}\right]\right)^{N-1}}\nonumber\\
&-& \left(1-\alpha\exp\left[-\frac{(x-0.5)^2}{W^2}\right]\right)^{M-N},
\end{eqnarray}
where again $x$ has been scaled to code units so that $x\in[0,1]$, $W$, $\alpha$ are numerical constants which respectively specify the width and height of the density field, which obviously peaks at $x=0.5$. Such a density field has been used in the code test of \cite{ecosmog}.

Note that such a density field is not exactly periodic at the edges of the simulation box, but given that $W$ is small enough, $\tilde{\rho}\rightarrow0$ at the box edges and periodic boundary conditions are approximately satisfied.

\begin{figure}
\includegraphics[scale=0.36]{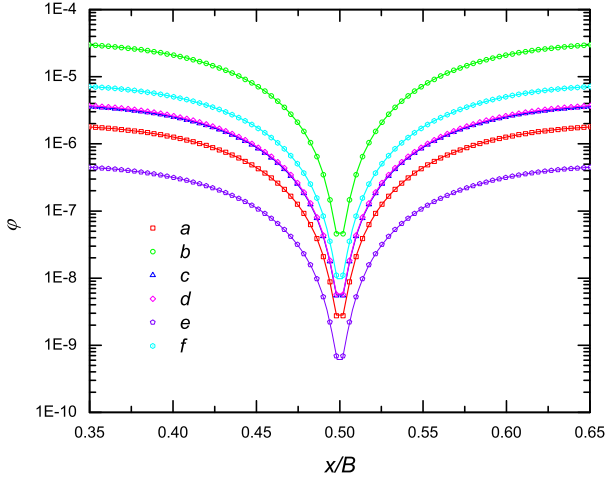}
\caption{(Colour online) Solutions of $\varphi$ in a one-dimensional ($x$-direction) Gaussian-type density field constructed using Eq.~(\ref{eq:gaussian_dens_symmetron}), for the six test symmetron models in Table~\ref{tab:test_models_symmetron} (see legends). The solid curves are the analytical results from Eq.~(\ref{eq:varphi_gaussian_dens_symmetron}) and the symbols with same colours are the corresponding numerical solutions. A simulation box with side length of $250h^{-1}$Mpc and 256 grid cells on each side is used in the computation and the symmetron equation is only solved on the regular domain grid. $x$ is rescaled so that $x/B\in[0,1]$.} \label{fig:test_gaussian_dens_symmetron}
\end{figure}

The solution to $\varphi$ can then be obtained analytically and is
\begin{eqnarray}\label{eq:varphi_gaussian_dens_symmetron}\label{eq:varphi_gaussian_dens_dilaton}
\varphi(x) &=& \BLED{\varphi_\star}\left[1-\alpha\exp\left(-\frac{(x-0.5)^2}{W^2}\right)\right],
\end{eqnarray}
which clearly shows that when $\alpha\rightarrow1$ $|\varphi|$ could be made very small at $x=0.5$ while at $x\rightarrow0$ or $x\rightarrow1$ it goes to $\varphi=\BLED{\varphi_\star}$.

We have implemented Eq.~(\ref{eq:gaussian_dens_symmetron}) into our numerical code and the numerical solutions for $\varphi$ are shown in Fig.~\ref{fig:test_gaussian_dens_symmetron}. We can see that they agree with the analytical solution Eq.~(\ref{eq:varphi_gaussian_dens_symmetron}) very well.

For the dilaton case we use the following density field
\begin{eqnarray}\label{eq:gaussian_dens_dilaton}
\tilde{\rho}(x) &=& \frac{\tilde{c}^2a}{3\Omega_mA_2}\frac{2\alpha}{W^2}\frac{\exp\left[-\frac{(x-0.5)^2}{W^2}\right]\left[1-2\frac{(x-0.5)^2}{W^2}\right]}{1-\alpha\exp\left[-\frac{(x-0.5)^2}{W^2}\right]}\\
&&+\left[a^{2r-3}+\frac{2r-3}{s}\log\left[1-\alpha e^{-\frac{(x-0.5)^2}{W^2}}\right]\right]^{-\frac{3}{2r-3}}a^3\nonumber
\end{eqnarray}
\BLED{where $x$, $W$ and $\alpha$ are specified similarly as above.}

\begin{figure}
\includegraphics[scale=0.36]{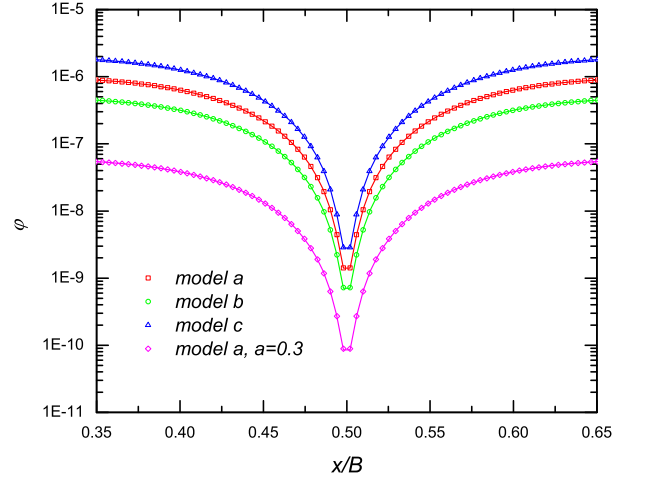}
\caption{(Colour online) Solutions of $\varphi$ in a one-dimensional ($x$-direction) Gaussian-type density field constructed using Eq.~(\ref{eq:gaussian_dens_dilaton}), for three test dilaton models (a, b, c) at $a=1.0$ and test model a at $a=0.3$ (see legends). The solid curves are the analytical predictions from Eq.~(\ref{eq:varphi_gaussian_dens_dilaton}) and the symbols with same colours are the corresponding numerical solutions. Other specifications are the same as in Fig.~\ref{fig:test_gaussian_dens_symmetron}.} \label{fig:test_gaussian_dens_dilaton}
\end{figure}

The test results for the dilaton models are shown in Fig.~\ref{fig:test_gaussian_dens_dilaton}, where again we find good agreement with the analytical solution Eq.~(\ref{eq:varphi_gaussian_dens_dilaton}).

\subsection{Equation solver on refinements}

\begin{figure}
\includegraphics[scale=0.36]{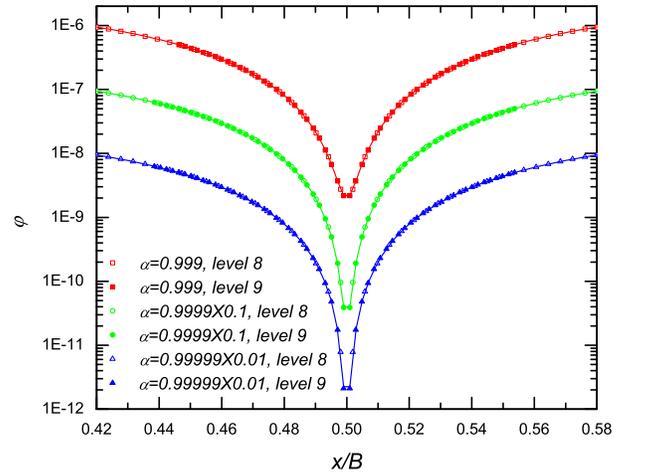}
\caption{(Colour online) Same as Fig.~\ref{fig:test_gaussian_dens_symmetron}, but for the model a only and $\alpha=0.999, 0.9999, 0.99999$ (from top to bottom: red, green, blue). The symmetron equation is solved on two levels: level 8 (the regular domain grid) and level 9 (the first refinement), and their numerical solutions are represented by empty and filled symbols of the same shape and colour respectively. The solid curves of the same colours are the corresponding analytical solutions from Eq.~(\ref{eq:varphi_gaussian_dens_symmetron}). A simulation box with side length of $250h^{-1}$Mpc and 256 grid cells on each side is used in the computation and the symmetron equation is only solved on the regular domain grid. $x$ is rescaled so that $x/B\in[0,1]$. For clarity we have multiplied the results for $\alpha=0.9999$ and $0.99999$ by $0.1$ and $0.01$ respectively.} \label{fig:test_multilevel_symmetron}
\end{figure}

The above tests show that our solver of the scalar field EOM works accurately on regular grids. But in cosmological simulations these equations are also solved on irregularly-shaped refinements where they can take different forms due to the refinement boundaries \cite{ecosmog}. It is therefore necessary to test the scalar field equation solver on refinements as well, which we will do in this subsection.

The Gaussian-type density configuration provides a good way to check the multilevel scalar-equation solver, because the density peak can be made arbitrarily high by adjusting the parameter $\alpha$ \HANS{and the value of the matter density is the criterion we use to refine grid cells in cosmological simulations.} In the vicinity of this peak, the density field $\tilde{\rho}$ changes rapidly and higher spatial resolution is necessary to compute $\varphi$ \BLED{(and differentiate it to get the fifth force)} accurately.

Consider the case where the regular domain grid is refined only once, in regions where the density value exceeds a given threshold (we call this a `two-level problem', and in the numerical examples below the coarse and fine levels are respectively levels 8 and 9). The density values $\tilde{\rho}$ in both the coarse and the refined cells are calculated using Eq.~(\ref{eq:gaussian_dens_symmetron}) for the symmetron case and Eq.~(\ref{eq:gaussian_dens_dilaton}) for the dilaton case, while the values of $\varphi$ at the fine-level boundaries are computed from interpolation of those in the nearby coarse-level cells \cite{ecosmog}.

Fig.~\ref{fig:test_multilevel_symmetron} shows the numerical values of $\varphi$ on both levels in the region covered by the refinement, for the symmetron case. We show the results for model a only and for four different values of $\alpha$ ($0.999$, $0.9999$ and $0.99999$ from top to bottom), and for each $\alpha$ the results from the coarse and fine levels are denoted respectively by empty and filled symbols. For comparison we have also plotted the analytical results Eq.~(\ref{eq:varphi_gaussian_dens_symmetron}) as solid curves. As we can see, both fine-level and coarse-level results are virtually indistinguishable from the exact solution.

This does not mean that the refinement is unnecessary however, because, as shown in Fig.~\ref{fig:test_multilevel_symmetron}, the fine level has more data points and could probe regions closer to the extreme value of $\varphi$, which corresponds to the high density region where high resolution is needed.

\begin{figure}
\includegraphics[scale=0.36]{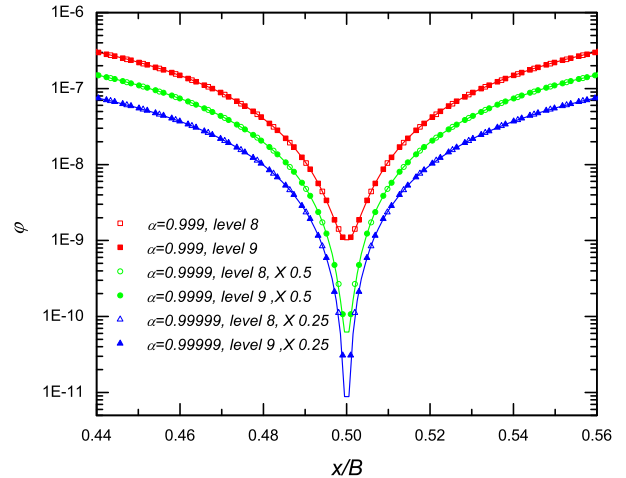}
\caption{(Colour online) Same as Fig.~\ref{fig:test_gaussian_dens_dilaton}, but for the model a only and $\alpha=0.999, 0.9999, 0.99999$ (from top to bottom: red, green, blue). The dilaton equation is solved on two levels: level 8 (the regular domain grid) and level 9 (the first refinement), and their numerical solutions are represented by empty and filled symbols of the same shape and colour respectively. The solid curves of the same colours are the corresponding analytical solutions from Eq.~(\ref{eq:varphi_gaussian_dens_dilaton}). A simulation box with side length of $250h^{-1}$Mpc and 256 grid cells on each side is used in the computation and the dilaton equation is only solved on the regular domain grid. $x$ is rescaled so that $x/B\in[0,1]$. For clarity we have multiplied the results for $\alpha=0.9999$ and $0.99999$ by $0.5$ and $0.25$ respectively.} \label{fig:test_multilevel_dilaton}
\end{figure}

For the dilaton, Fig.~\ref{fig:test_multilevel_dilaton} shows the numerical values of $\varphi$ on both levels in the region covered by the refinement. Again, we show the results for model a only and for four different values of $\alpha$ ($0.999$, $0.9999$ and $0.99999$ from top to bottom), and for each $\alpha$ the results from the coarse and fine levels are denoted respectively by empty and filled symbols.
 For comparison we have also plotted the analytical results Eq.~(\ref{eq:varphi_gaussian_dens_dilaton}) as solid curves. Excellent agreement is found again.

\subsection{Other tests}

\BLED{In the above we have focused on various tests of the scalar field solver of the {\tt ECOSMOG} code, as this is the only new addition to the default {\tt RAMSES} $N$-body code. These tests checked the validity of the new subroutines against different density distributions, and the good agreements with analytical solutions shows the validity of the code and its accuracy.}

\BLED{As the standard gravity solver and particle-updating subroutines of {\tt RAMSES} are not touched, tests carried out for them (which show that the {\tt RAMSES} code works very well) need not be repeated here. The AMR simulation algorithm is often implemented in different ways in different codes; for a detailed explanation of its implementation in {\tt RAMSES} and therefore in {\tt ECOSMOG} we  refer to \citep{ramses} and \citep{ecosmog} respectively. We do not present the full details here as they are too long and  this paper is mainly concerned with the modified gravity physics.}

\BLED{When a new code is written, one needs to test its cosmological simulations. This is straightforward for a standard code of $\Lambda$CDM simulations, because there are fitting formulae and results from other codes to compare to. Unfortunately, up to now there are no  accurate fitting formulae for modified gravity theories such as symmetron, dilaton and $f(R)$ gravity. But several serial $N$-body codes simulating $f(R)$ gravity (e.g., \citep{o2008,zlk2011}) and symmetron models (e.g., \citep{dlmw2012}) do exist in the literature: in both cases good agreement with {\tt ECOSMOG} has been found\footnote{Another independent code which is still being developed also agrees with {\tt ECOSMOG} very well.}. See, for example, \citep{ecosmog} for a comparison for $f(R)$ gravity, and we have also checked explicitly that our symmetron simulation result agrees with that of \citep{dlmw2012}.} 

\BLED{Finally, for cases where approximate analytical results can be obtained from other methods, we find good agreement between {\tt ECOSMOG} and the approximation solutions. An example is the $f(R)$ gravity model of \citep{hs2007} with $|\rd f/\rd R|=10^{-4}$, the nonlinearity of which is very weak and so the matter power spectrum can be approximated by linear perturbation theory down to relatively small scales. This is actually confirmed in \citep{lhkzjb2012}, which can serve as another test of the {\tt ECOSMOG} code.}

\BLED{In short, the {\tt ECOSMOG} scalar field solver has been tested in various ways, and several cosmological simulations of modified gravity models using {\tt ECOSMOG} agree with similar simulations done using other codes, such as the codes developed independently in \citep{o2008,zlk2011,dlmw2012}.}

\section{Cosmological Simulations}

\label{sect:sim}

In this section we describe and analyse the results of cosmological simulations of the dilaton and symmetron modified gravity models. We also perform $\Lambda$CDM simulations for comparison. For each model we run 5 realisations with the same physical parameters and simulation specification, but different realisations of initial conditions. The initial conditions are generated using {\tt MPGRAFIC} \citep{mpgrafic} at redshift $z_i=49.0$ with different seeds of random numbers. Since at $z_i=49.0$ the effect of the fifth force is negligible, the initial conditions should be the same for all models studied here. For the ease of comparison, we use the same random seed to generate initial conditions for the same realisation of all models, including symmetron, dilaton and $\Lambda$CDM.

The background expansion history in the studied dilaton and symmetron models is in practice indistinguishable from that of the fiducial $\Lambda$CDM model \citep{bdlw2012}. In all simulations we adopt WMAP7 \cite{wmap7} cosmological parameters, with $h=0.71$, $\Omega_m=0.267$, $\Omega_\Lambda=0.733$, $n_s=0.963$ and $\sigma_8=0.801$.

The size of the simulation box is chosen to be 128$h^{-1}$Mpc, and the domain grid\footnote{As {\tt RAMSES} and {\tt ECOSMOG} are adaptive mesh refinements codes, the domain grid is defined as the finest uniform (regular) grid which covers the whole simulation box.} has $2^8=256$ cells on each side. The grid cells are refine  when the effective number of particles in them exceeds 9.0, and the finest refinement level equivalently has $2^{14}$ cells on each side. The number of particles is $N_p = 256^3$ in all simulations.

\begin{figure*}
\includegraphics[scale=0.6]{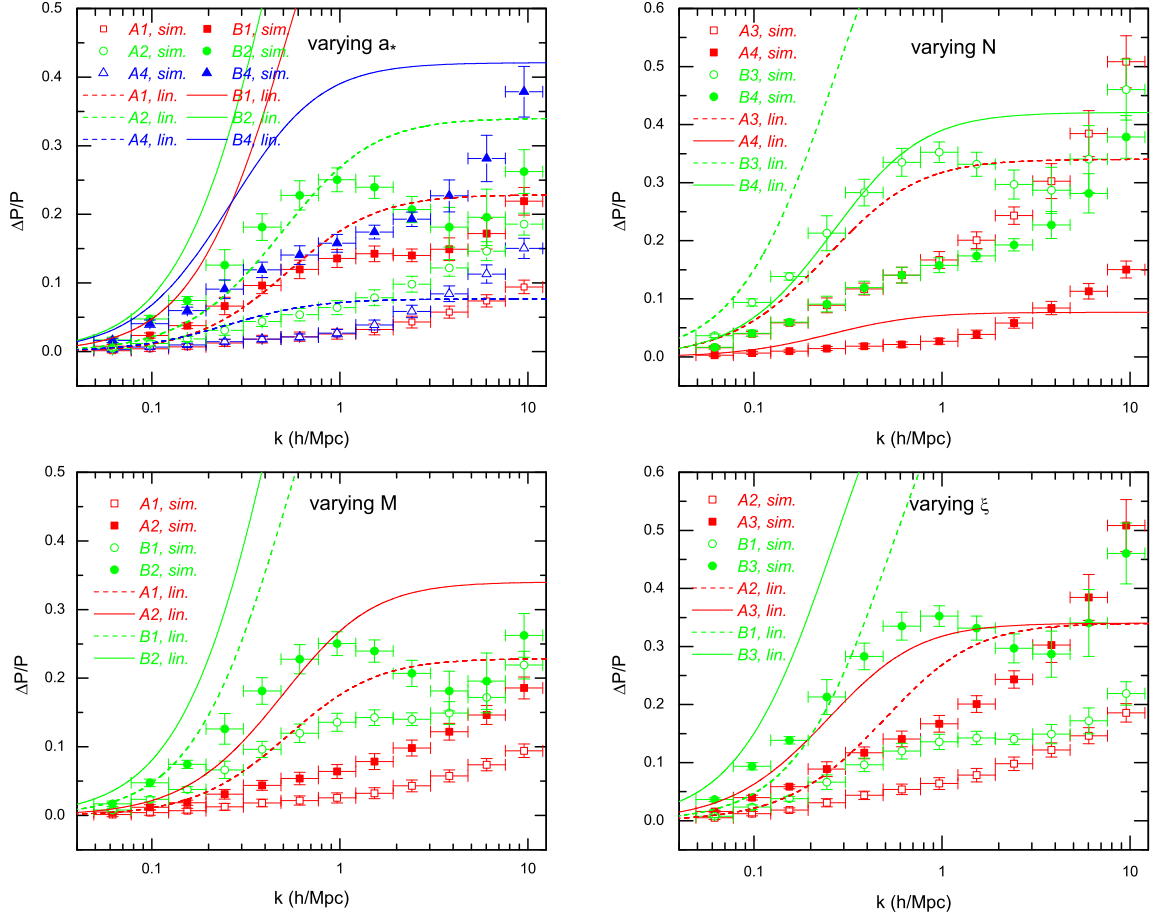}
\caption{(Colour online) The relative difference between the matter power spectra of the symmetron models and the $\Lambda$CDM paradigm. The symbols are from the $N$-body simulations, and the curves are linear perturbation theory predictions. Details are illustrated by the legends, and $a=1.0$.}
\label{fig:dpop_a_1.0_sym}
\end{figure*}

\begin{figure*}
\includegraphics[scale=0.6]{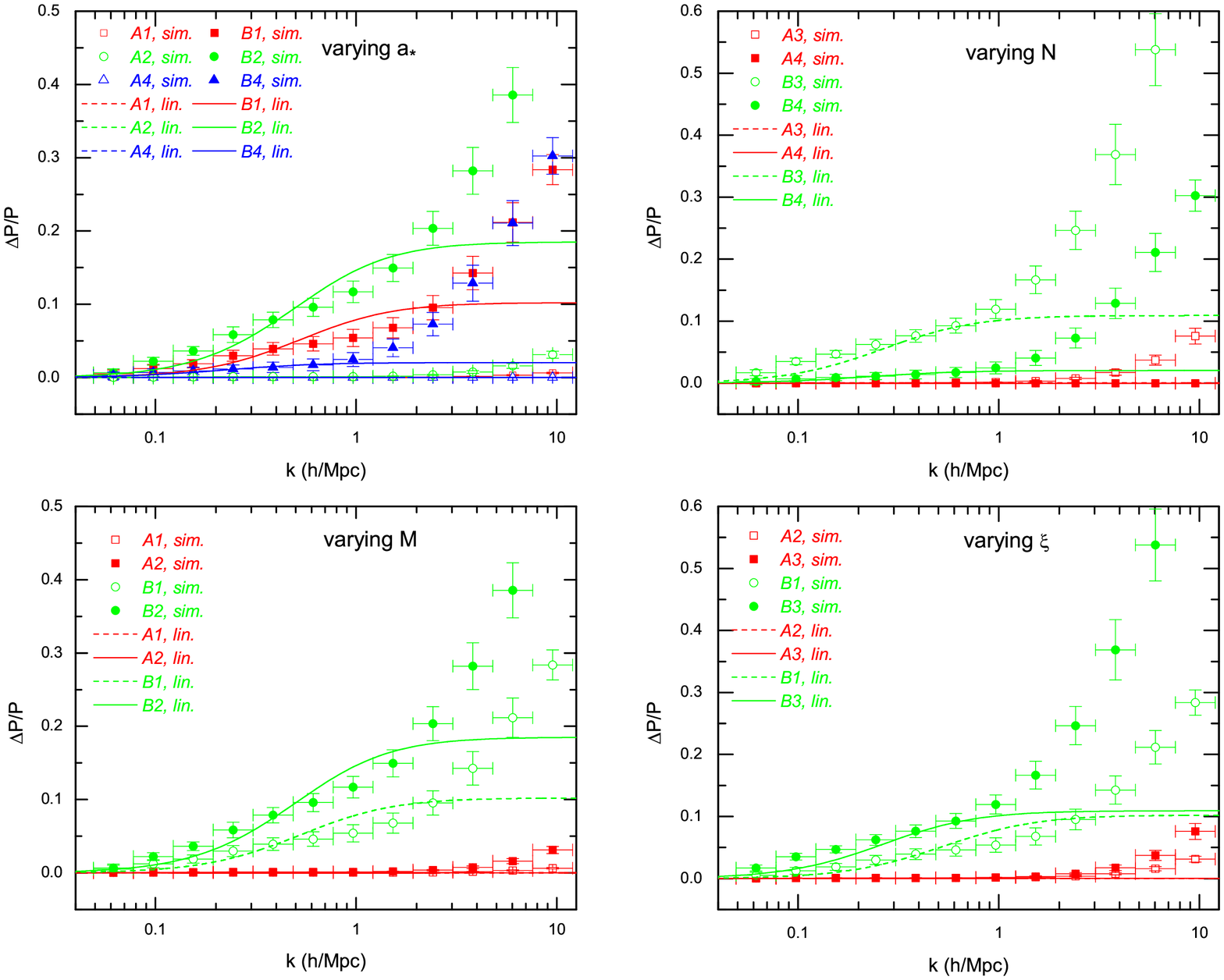}
\caption{(Colour online) The same as Fig.~\ref{fig:dpop_a_1.0_sym}, but for $a=0.5$.}
\label{fig:dpop_a_0.5_sym}
\end{figure*}


\begin{figure*}
\includegraphics[scale=0.6]{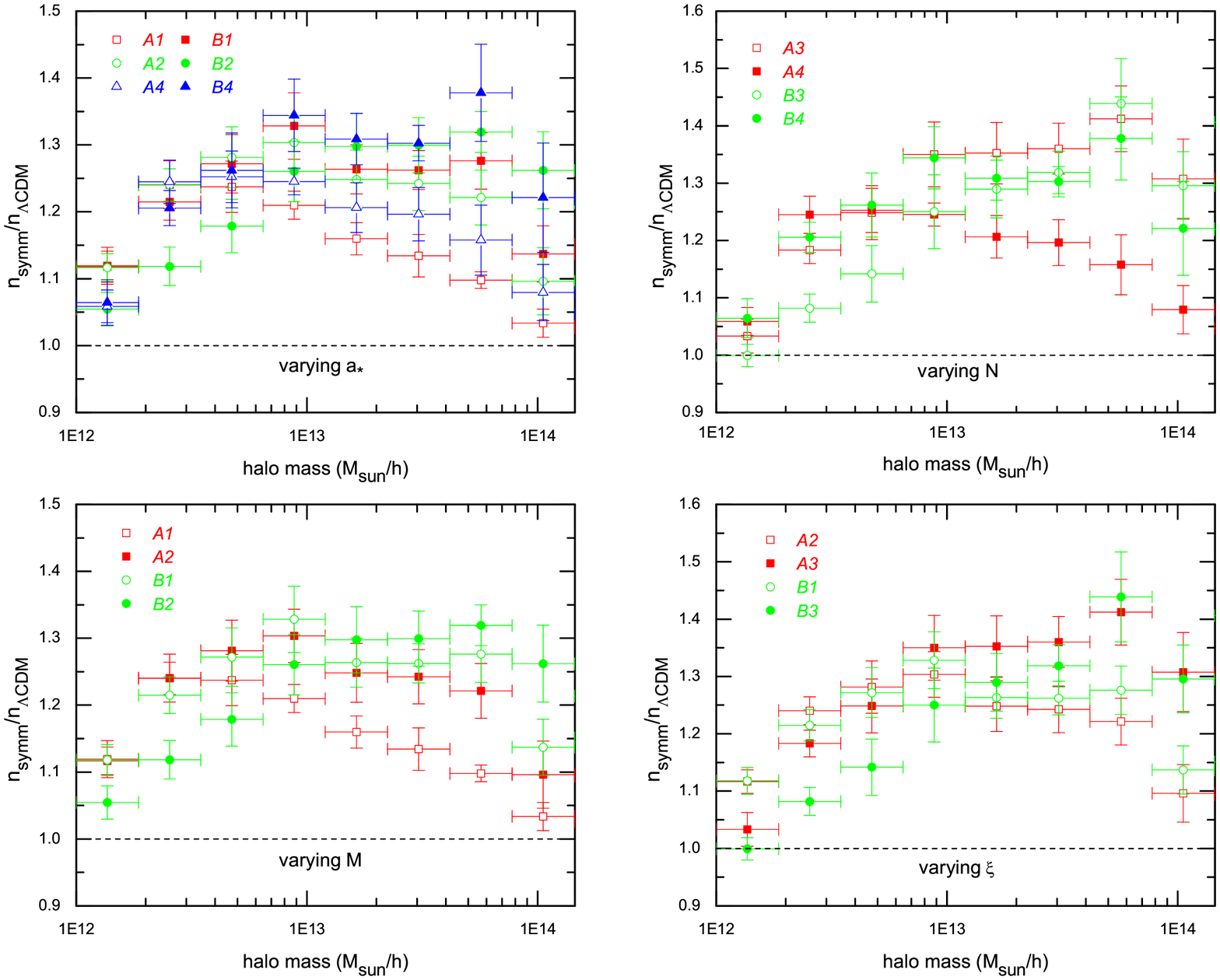}
\caption{(Colour online) The ratio between the mass functions of the symmetron models and the $\Lambda$CDM paradigm at $a=1.0$.}
\label{fig:dnon_a_1.0_sym}
\end{figure*}

\begin{figure*}
\includegraphics[scale=0.6]{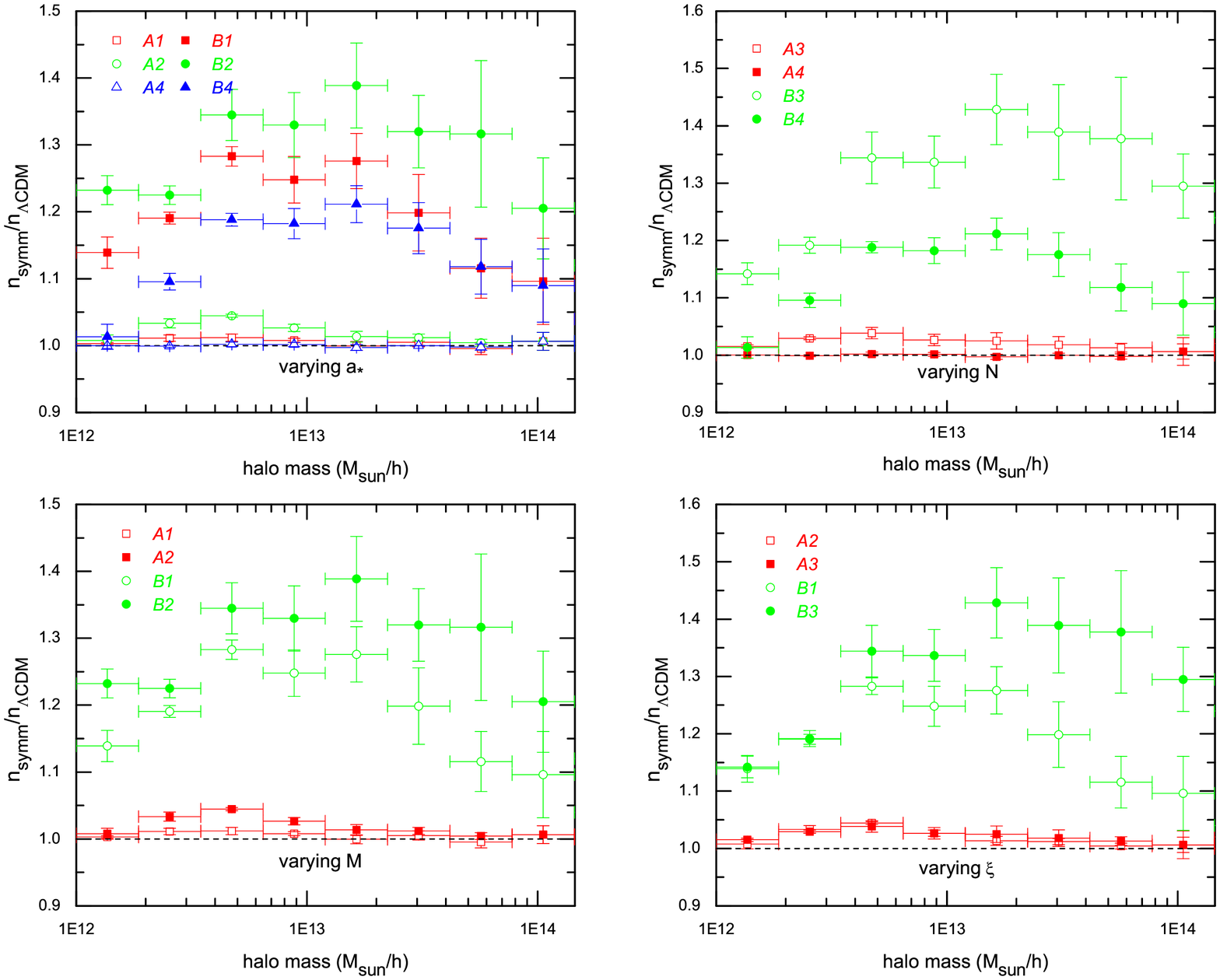}
\caption{(Colour online) The same as Fig.~\ref{fig:dnon_a_1.0_sym}, but for $a=0.5$.}
\label{fig:dnon_a_0.5_sym}
\end{figure*}


\subsection{The symmetron models}

\label{subsect:sim_symmetron}

The symmetron models are specified by the four model parameters $a_\ast$, $M$, $N$ and $\xi$. We have chosen to fix $\BLED{\beta_\star}=1.0$ for all our runs in order to see the effect of varying the other parameters individually. The effect of varying $\BLED{\beta_\star}$ is to modulate the strength of the fifth force and was investigated for the symmetron in \cite{dlmw2012}. In Table~(\ref{tab:run_models_symmetron}) we list the parameters for the nine models we have simulated.

\BLED{In the rest of this subsection, we will focus on the effects of changing each model parameter on the major cosmological observables such as the matter power spectrum and halo mass function.
More specifically, we will analyse the results of our numerical simulations according to the following:}
\begin{enumerate}
\item How the \HANS{symmetry breaking} scale factor $a_ \ast$ affects the results: Model A1 versus B1, A2 versus B2 and A4 versus B4.
\item How the coupling strength parameter $N$ affects the results: Model A3 versus A4 and B3 versus B4.
\item How the potential parameter $M$ influences the results: Model A1 versus A2 and B1 versus B2.
\item How the range $\BLED{\lambda_\star} \equiv 2998\xi$ Mpc$/h$ of the fifth force influences the results: Model A2 versus A3 and B1 versus B3.
\end{enumerate}

\begin{table}
\caption{The parameter values for the nine models used in the symmetron cosmological simulations. For each model we have 5 realisations of initial conditions, and therefore a total of 45 runs.}
\begin{tabular}{@{}lccccc}
\hline\hline
model name\ \ \ \  & $a_\ast$\ \ \ \  & $\BLED{\beta_\star}$\ \ \ \  & $(N,M)$\ \ \ \  & $2998\xi$\ \ \ \ \ & realisations\\
\hline
$\Lambda$CDM\ \ \ \  & $-$\ \ \ \ &  $-$\ \ \ \  & $-$\ \ \ \  & $-$\ \ \ \ & 5\\
 \ \ \ \  & $$\ \ \ \  & $$\ \ \ \  & $$\ \ \ \  & $$\ \ \ \ & \\
A1\ \ \ \  & $0.50$\ \ \ \ &  $1.0$\ \ \ \  & $(2,4)$\ \ \ \  & $1.0$\ \ \ \ &5\\
A2\ \ \ \  & $0.50$\ \ \ \  & $1.0$\ \ \ \  & $(2,6)$\ \ \ \  & $1.0$\ \ \ \ &5\\
A3\ \ \ \  & $0.50$\ \ \ \  & $1.0$\ \ \ \  & $(2,6)$\ \ \ \  & $2.0$\ \ \ \ &5\\
A4\ \ \ \  & $0.50$\ \ \ \  & $1.0$\ \ \ \  & $(4,6)$\ \ \ \  & $2.0$\ \ \ \ &5\\
 \ \ \ \  & $$\ \ \ \  & $$\ \ \ \  & $$\ \ \ \  & $$\ \ \ \ & \\
B1\ \ \ \  & $0.33$\ \ \ \  & $1.0$\ \ \ \  & $(2,4)$\ \ \ \  & $1.0$\ \ \ \ &5\\
B2\ \ \ \  & $0.33$\ \ \ \  & $1.0$\ \ \ \  & $(2,6)$\ \ \ \  & $1.0$\ \ \ \ &5\\
B3\ \ \ \  & $0.33$\ \ \ \  & $1.0$\ \ \ \  & $(2,4)$\ \ \ \  & $2.0$\ \ \ \ &5\\
B4\ \ \ \  & $0.33$\ \ \ \  & $1.0$\ \ \ \  & $(4,6)$\ \ \ \  & $2.0$\ \ \ \ &5 \\
\hline\hline
\end{tabular}\label{tab:run_models_symmetron}
\end{table}

\subsubsection{Nonlinear matter power spectra}

The most direct way to see the effect of modified gravity on the clustering of matter is to look at the matter power spectrum $P(k)$. We have measured the nonlinear $P(k)$ in the symmetron models and calculated their relative differences from the $\Lambda$CDM prediction. The results are shown in Figs.~\ref{fig:dpop_a_1.0_sym}, \ref{fig:dpop_a_0.5_sym}. The power spectra are measured using the publicly available code {\tt POWMES} \cite{powmes}.

\begin{enumerate}
\item The \BLED{\g{symmetry} breaking} scale factor $a_ \ast$ controls when the fifth force starts to kick in. From Fig.~\ref{fig:dpop_a_1.0_sym} we could see that decreasing $a_\ast$ (i.e., moving from A models to B models) leads to a stronger matter power spectrum as the fifth force would have more time to participate in structure formation. Notice that when $a\leq a_\ast$ the matter power spectra in symmetron models are essentially unchanged as can be seen in Fig.~\ref{fig:dpop_a_0.5_sym}\footnote{In Fig.~\ref{fig:dpop_a_0.5_sym} symmetry breaking has just happened for A models and the fifth-force effect has not accumulated at $a=0.5$.}. This is because on linear scales there is strictly no fifth-force effect before $a=a_\ast$, since the magnitude of the fifth force is determined by the {\it background} matter density, which is always higher than $\rho_\ast$ before $a=a_\ast$. However, on nonlinear scales, the fifth force can kick in even before $a=a_\ast$ in regions where matter density drops below $\rho_\ast$, thus the structure formation is affected even at $a_\ast$.
\item The parameter $N$ of the matter coupling $\BLED{\beta\propto\varphi}^{N-1}$ determines how the matter coupling evolves. As the field moves towards $\BLED{\varphi}=0$ in high density regions, a larger $N$ means that the fifth force becomes more suppressed as shown in \cite{bdlw2012}. This effect can be seen in Fig.~\ref{fig:dpop_a_1.0_sym} (upper right panel). \BLED{Note that varying $N$ also changes the evolution of $\varphi$ through the changes of $\beta(\varphi)$ and $V(\varphi)$; however the numerical result here shows that this effect is subdominant.}
\item The parameter $M$ of the self-interaction term $\BLED{\varphi}^M \in V(\varphi)$ determines how \g{nonlinearly} the model behaves. A higher-order (larger $M$) interaction term means that the nonlinearities, and therefore the screening mechanism, are less at play (see \S~\ref{subsect:symmetron_effect}), which again leads to more matter clustering as confirmed by the lower-left panel of Fig.~\ref{fig:dpop_a_1.0_sym}.
This effect can also be seen by noting that the nonlinear power spectra for the cases of $M=6$ are in general closer to the corresponding linear power spectra than for the cases of $M=4$.
\item The range $\BLED{\lambda_\star} = 2998 \xi {\rm Mpc}/h$ of the fifth force determines which scales  are influenced by the fifth force. Increasing the range moves the modifications of gravity to larger cosmological scales as can be seen in Fig.~\ref{fig:dpop_a_1.0_sym}. In the linear perturbation regime, the power spectra for two models with different ranges ($\BLED{\lambda_{\star1,2}}$) are related by \g{the scaling relation} $P_1(k) = P_2(k \BLED{\lambda_{\star1}/\lambda_{\star2})}$. \g{However, this scaling} no longer holds in the nonlinear regime. \g{For example, when $\BLED{\lambda_\star}$ decreases,} the symmetron mass becomes heavier, the screening effect is enhanced and consequently the power spectrum is suppressed (c.f.~Fig.~\ref{fig:dpop_a_1.0_sym} and \S~\ref{subsect:symmetron_effect}).
\item \BLED{At late times (Fig.~\ref{fig:dpop_a_1.0_sym}) the linear perturbation prediction is a bad approximation to the full solution, which is because the symmetron EOM is highly nonlinear. Indeed, as in the case of $f(R)$ gravity \citep{lhkzjb2012}, the linear theory becomes inaccurate almost as soon as the power spectrum starts to deviate from the $\Lambda$CDM prediction. This shows the importance of properly taking into account the nonlinear effects (by numerical simulations) in the study of structure formation in modified gravity models.}
\item \BLED{The agreement between the linear and nonlinear results becomes better at earlier times (c.f.~Fig.~\ref{fig:dpop_a_0.5_sym}), when the effect of nonlinearity has not accumulated for long.}
\end{enumerate}


\BLED{In $f(R)$ gravity models, it is known \citep{lhkzjb2012} that the shape of $\Delta P/P$ follows a fixed evolution path, and at any given time the position of a model on this path is determined by the properties of the fifth force and how long it has become non-negligible. Similar patterns appear here, for example in the A models $a_\ast=0.5$ where the fifth force becomes non-negligible later than in the B models, for which $a_\ast=0.3$. Correspondingly, in Fig.~\ref{fig:dpop_a_1.0_sym} $\Delta P/P$ has a peak at $k\sim1h$Mpc$^{-1}$. On the other hand, Fig.~\ref{fig:dpop_a_1.0_sym} shows that for symmetron models $\Delta P/P$ goes up again on very small scales ($k\geq{\rm a~few}$), while in $f(R)$ models $\Delta P/P$ decreases for these scales \citep{lhkzjb2012}.}




\g{This pattern for the symmetron matter power spectrum can be understood as follows.
At early times the model is well described by linear perturbation theory and the symmetron mass (and the coupling strength $\beta(\varphi)$) is nearly the same everywhere; the Yukawa nature necessarily means that the fifth force decays with distance, and as a result} $\Delta P/P$ increases monotonically with $k$ at these times (see Fig.~\ref{fig:dpop_a_0.5_sym}). Later when highly nonlinear and dense structures have formed, the symmetron screening mechanism starts to work so that the fifth force inside these structures are efficiently suppressed ($\beta(\varphi)$ becomes small) and GR is locally restored since then, which makes $\Delta P/P$ frozen on small scales (thus remain monotonically increasing) while at the same time still grow on larger scales (e.g., $k\gtrsim1h$Mpc$^{-1}$) as the fifth force still propagates among different halos.

\BLED{To understand this behaviour more properly would require a detailed study of the density and velocity fields, together with their time evolutions, and these will be left to future work with higher-resolution and larger simulations.}

As an illustration of the above effects, the difference between the symmetron models we have simulated and $\Lambda$CDM on scales of order 1 Mpc can be as large as 30 percent today. This can be seen
in Fig.13 for models B1 and B3 where the range of the force is respectively 1 and 2 Mpc and the highest power in the potential is 6 and 4 respectively. On these examples, the characteristic bump of the symmetron models can also be seen in a clear way.


\subsubsection{Mass functions}

We have measured the mass functions from our simulations using the publicly available code {\tt AHF} \cite{ahf}, which is efficiently parallelised using {\tt MPI} and {\tt OpenMP}. The mass of a halo is defined as the total mass contained in $R_{200}$, the radius at which the density contrast $\Delta$ drops below 200 times the critical density. For each model, including $\Lambda$CDM, we have calculated the average and standard deviation of the mass function over the five realisations.

Because we are interested in how the fifth force can change the matter clustering, we show the ratio of the symmetron and $\Lambda$CDM mass functions, $\BLED{\mathcal{R}}\equiv n_{\rm symmetron}/n_{\rm\Lambda CDM}$. The standard deviation $\sigma_\mathcal{R}$ of $\BLED{\mathcal{R}}$ for each mass bin is computed using the normal rule of propagation of errors, according to which we have
\begin{eqnarray}
\left(\frac{\sigma_{\BLED{\mathcal{R}}}}{\BLED{\mathcal{R}}}\right)^2 &=& \left(\frac{\sigma_{\rm MG}}{n_{\rm MG}}\right)^2 + \left(\frac{\sigma_{\rm\Lambda}}{n_{\rm\Lambda}}\right)^2-2\BLED{\hat{\rho}}\frac{\sigma_{\rm MG}}{n_{\rm MG}}\frac{\sigma_{\rm \Lambda}}{n_{\rm\Lambda}},\
\end{eqnarray}
The subscripts $_{\rm MG}$ and $_{\rm\Lambda}$ denote the modified gravity model (the symmetron here and the dilation in the next section) and $\Lambda$CDM respectively, and $\BLED{\hat{\rho}}$ is the correlation coefficient between the mass functions of the two, i.e.,

\begin{equation}
\BLED{\hat{\rho}}=\frac{\sum\limits_{i}\left(n_{\rm MG}^i-\bar{n}_{\rm MG}\right)\left(n_{\Lambda}^i-\bar{n}_{\Lambda}\right)}{\left[\sum\limits_{i}\left(n_{\rm MG}^i-\bar{n}_{\rm MG}\right)^2\sum\limits_{i}\left(n_{\Lambda}^i-\bar{n}_{\Lambda}\right)^2\right]^{1/2}}
\end{equation} where the sum is over five realisations and the quantity with an overbar denotes the average over five realisations.

In Fig.~\ref{fig:dnon_a_1.0_sym} we show the ratios between the symmetron and $\Lambda$CDM mass functions from our simulations at $a=1.0$. The results at $a=0.5$ are shown in Fig.~\ref{fig:dnon_a_0.5_sym}.

The fifth force leads to an overall enhancement of the formation of dark matter structures. The effect is strongest for intermedium-sized ($M \sim 10^{13}h^{-1}M_{\odot}$) halos and we find a maximum enhancement in the mass function of around $50\%$ compared to $\Lambda$CDM for the models we have simulated. For the largest halo masses ($M \gtrsim 10^{14}h^{-1}M_{\odot}$) the symmetron mass function goes towards $\Lambda$CDM as the symmetron screening mechanism makes sure the fifth force is effectively suppressed for such massive objects.

The effects of varying different model parameters on the mass function are not as clear as in the power spectrum, but we can see the same trends. More specifically,
\begin{enumerate}
 \item For models with smaller $a_\ast$ (i.e., the B models) we see from Fig.~\ref{fig:dnon_a_1.0_sym} that a larger fraction of high mass halos is obtained. As with the matter power spectrum, the mass function is essentially unmodified for $a\leq a_\ast$ (see A models in Fig.~\ref{fig:dnon_a_0.5_sym}, for which $a_\ast=0.5$ and the effect of the fifth force has not accumulated at $a=0.5$). These are to be expected since the fifth force is not at play on cosmological scales at such early times, \BLED{and for smaller $a_\ast$ the fifth force has acted for a longer period. Hence more large halos form and fewer small halo survive the mergers}.
 \item \BLED{As mentioned in \S~\ref{subsect:symmetron_effect}, increasing the parameter $N$ leads to a suppression of the fifth force, especially for large halos and in high density regions where $|\varphi|\ll\varphi_\star$. This can be seen from the upper-right panel of Fig.~\ref{fig:dnon_a_1.0_sym}. Note that in models B3 and B4 both $N$ and $M$ are different, and the effect is not purely due to varying $N$.}
 \item \BLED{As discussed in \S~\ref{subsect:symmetron_effect}, increasing $M$ makes it easier for the scalar field to roll away from $\varphi=0$ where the coupling strength vanishes. This leads to a stronger fifth force and consequently more large halos, as can be seen in Fig.~\ref{fig:dnon_a_0.5_sym} (lower-left panel).}
\item Increasing $\xi$ increases the range $\BLED{\lambda_\star}$ of the fifth force and leads to more high-mass halos. This can be seen in Fig.~\ref{fig:dnon_a_1.0_sym}.
\end{enumerate}

\BLED{As for $\Delta P/P$, the effects of varying different model parameters on the shape of $\Delta n/n$ are similar, which shows that the four parameters are highly degenerate. This behaviour is different from what we will see in the dilaton simulations below.}

\BLED{The significant deviations of our symmetron models from the prediction of the $\Lambda$CDM paradigm, as shown in Figs.~\ref{fig:dnon_a_1.0_sym} and \ref{fig:dnon_a_0.5_sym}, should be detectable by future surveys.}

\subsection{The dilaton models}

\label{subsect:sim_dilaton}

In this subsection we analyse cosmological simulations of the generalised dilaton models. We vary all four model parameters $A_2, \beta_0, r$ and $\xi$, so that each of them takes 4 (3 for $A_2$) different values with the rest remaining the same. This results in a total of 12 dilaton models, as summarised in Table~\ref{tab:run_models_dilaton}. The choices of parameter values are such that A2, B2, C2 and D2 are the same model, to facilitate a cross comparison.

\BLED{As the dilaton simulations were run on a different machine from the symmetron ones, we have simulated the same default $\Lambda$CDM models on both machines, and checked that they agree very well. This enables a direct comparison between dilaton and symmetron simulations if needed.}

\begin{table}
\caption{The parameter values for the 65 cosmological simulations we have performed for this study. Note that '--' means that the parameters are unused for the $\Lambda$CDM case, and it means that the parameters are the same as in A2 in the cases of B2, C2 and D2.}
\begin{tabular}{@{}lcccccc}
\hline\hline
model name & $A_2$ & $\beta_0$ & $r$  & $\xi$ & realisations\\
\hline
$\Lambda$CDM\ \ & --\ \ & --\ \  & --\ \  & --\ \ & $5$\\
A1\ \  & $2.5\times10^5$\ \   & $0.50$\ \   & $1.00$\ \ & $0.001$\ \ & $5$\\
A2\ \  & $1.0\times10^5$\ \   & $0.50$\ \   & $1.00$\ \   & $0.001$\ \  & $5$\\
A3\ \  & $0.5\times10^5$\ \   & $0.50$\ \   & $1.00$\ \   & $0.001$\ \  & $5$\\
B1\ \  & $1.0\times10^5$\ \   & $0.25$\ \   & $1.00$\ \   & $0.001$\ \  & $5$\\
B2\ \ & --\ \ & --\ \  & --\ \  & --\ \ & $5$\\
B3\ \  & $1.0\times10^5$\ \   & $0.75$\ \   & $1.00$\ \   & $0.001$\ \  & $5$\\
B4\ \  & $1.0\times10^5$\ \   & $1.00$\ \   & $1.00$\ \   & $0.001$\ \  & $5$\\
C1\ \  & $1.0\times10^5$\ \   & \BLED{$0.50$}\ \  & $1.33$\ \   & $0.001$\ \  & $5$\\
C2\ \ & --\ \ & --\ \  & --\ \  & --\ \ & $5$\\
C3\ \  & $1.0\times10^5$\ \   & \BLED{$0.50$}\ \   & $0.67$\ \   & $0.001$\ \  & $5$\\
C4\ \  & $1.0\times10^5$\ \   & \BLED{$0.50$}\ \   & $0.40$\ \   & $0.001$\ \  & $5$\\
D1\ \  & $1.0\times10^5$\ \   & \BLED{$0.50$}\ \   & $1.00$\ \   & $0.0005$\ \  & $5$\\
D2\ \ & --\ \ & --\ \  & --\ \  & --\ \ & $5$\\
D3\ \  & $1.0\times10^5$\ \   & \BLED{$0.50$}\ \   & $1.00$\ \   & $0.002$\ \  & $5$\\
D4\ \  & $1.0\times10^5$\ \   & \BLED{$0.50$}\ \   & $1.00$\ \   & $0.003$\ \  & $5$\\
\hline\hline
\end{tabular}\label{tab:run_models_dilaton}
\end{table}

\begin{figure*}
\includegraphics[scale=0.6]{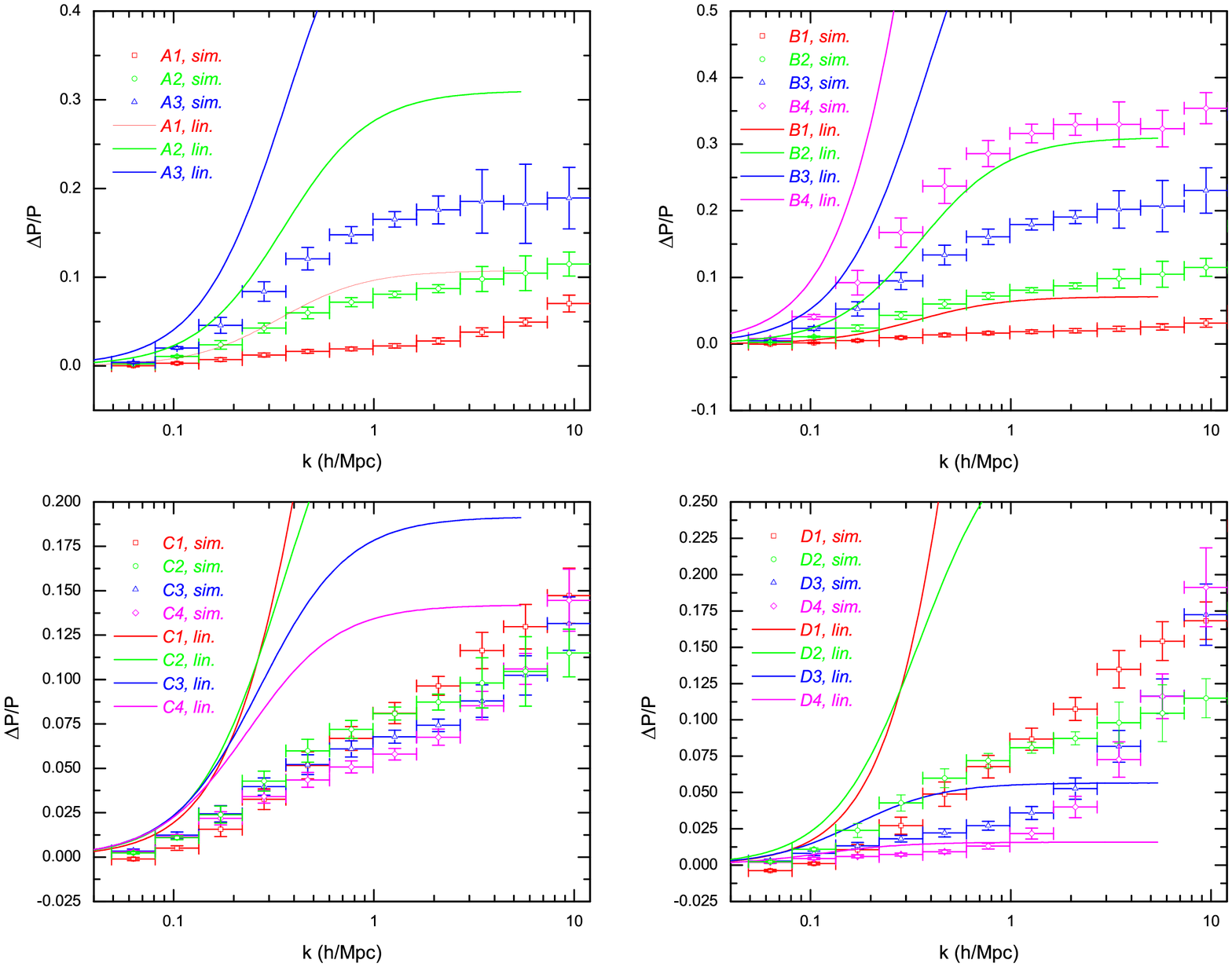}
\caption{(Colour online) The relative difference between the matter power spectra of the dilaton models and the $\Lambda$CDM paradigm. The symbols are from the $N$-body simulations, and the curves are linear perturbation theory predictions. Details are illustrated by the legends, and $a=1.0$.} \label{fig:dpop_a_1.0}
\end{figure*}

\begin{figure*}
\includegraphics[scale=0.6]{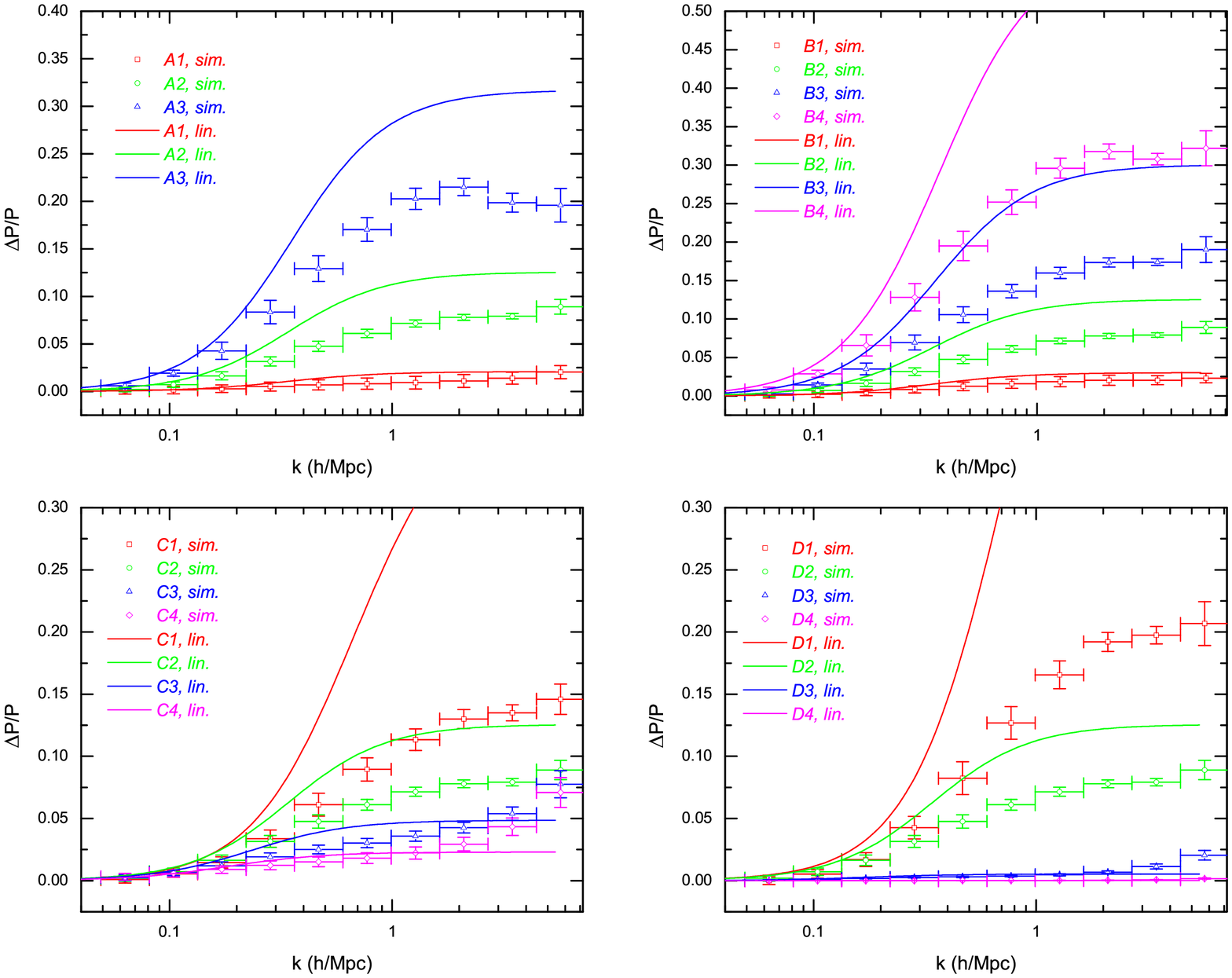}
\caption{(Colour online) The same as Fig.~\ref{fig:dpop_a_1.0}, but for $a=0.5$.} \label{fig:dpop_a_0.5}
\end{figure*}

\begin{figure*}
\includegraphics[scale=0.6]{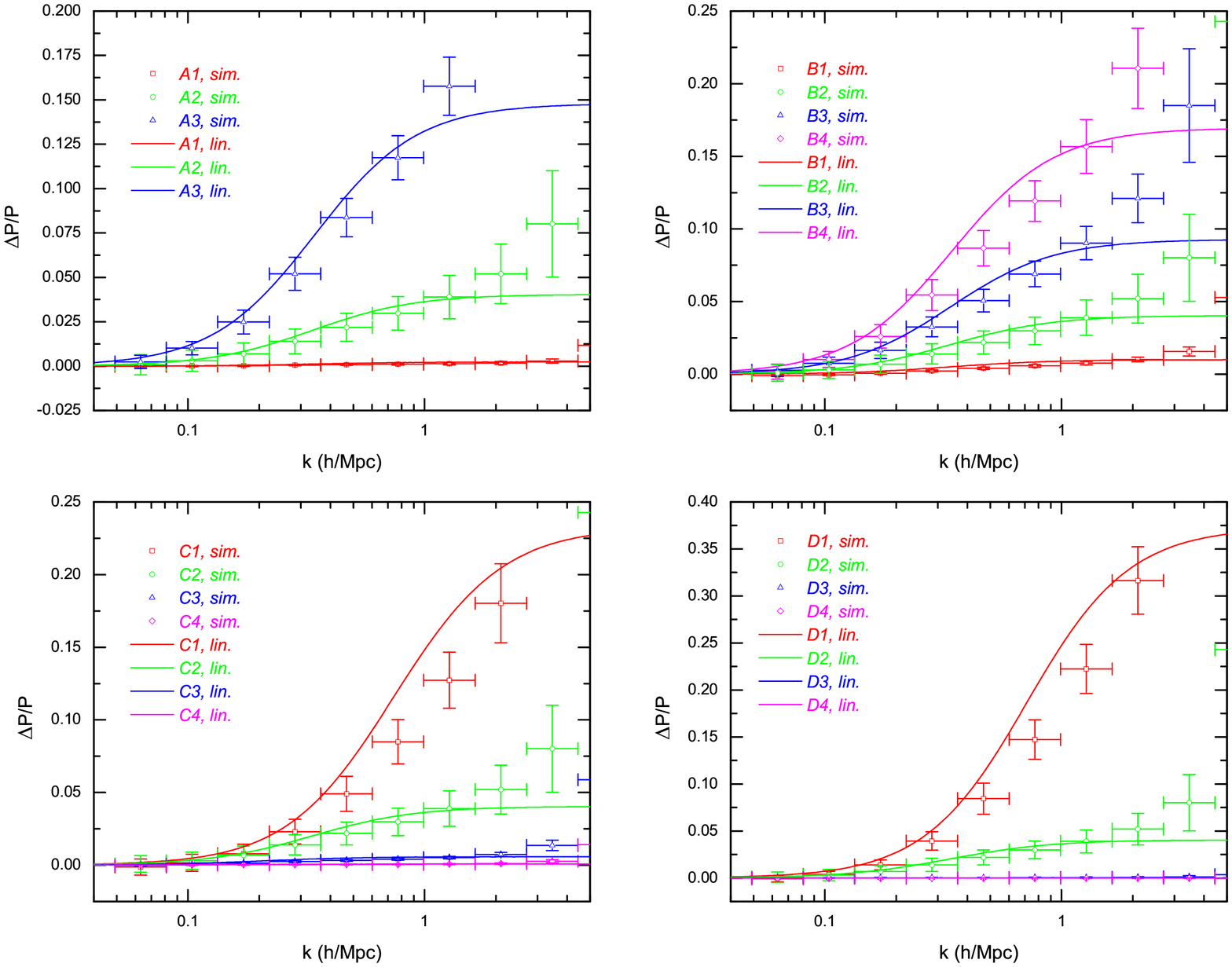}
\caption{(Colour online) The same as Fig.~\ref{fig:dpop_a_1.0}, but for $a=0.3$.} \label{fig:dpop_a_0.3}
\end{figure*}

\begin{figure*}
\includegraphics[scale=0.6]{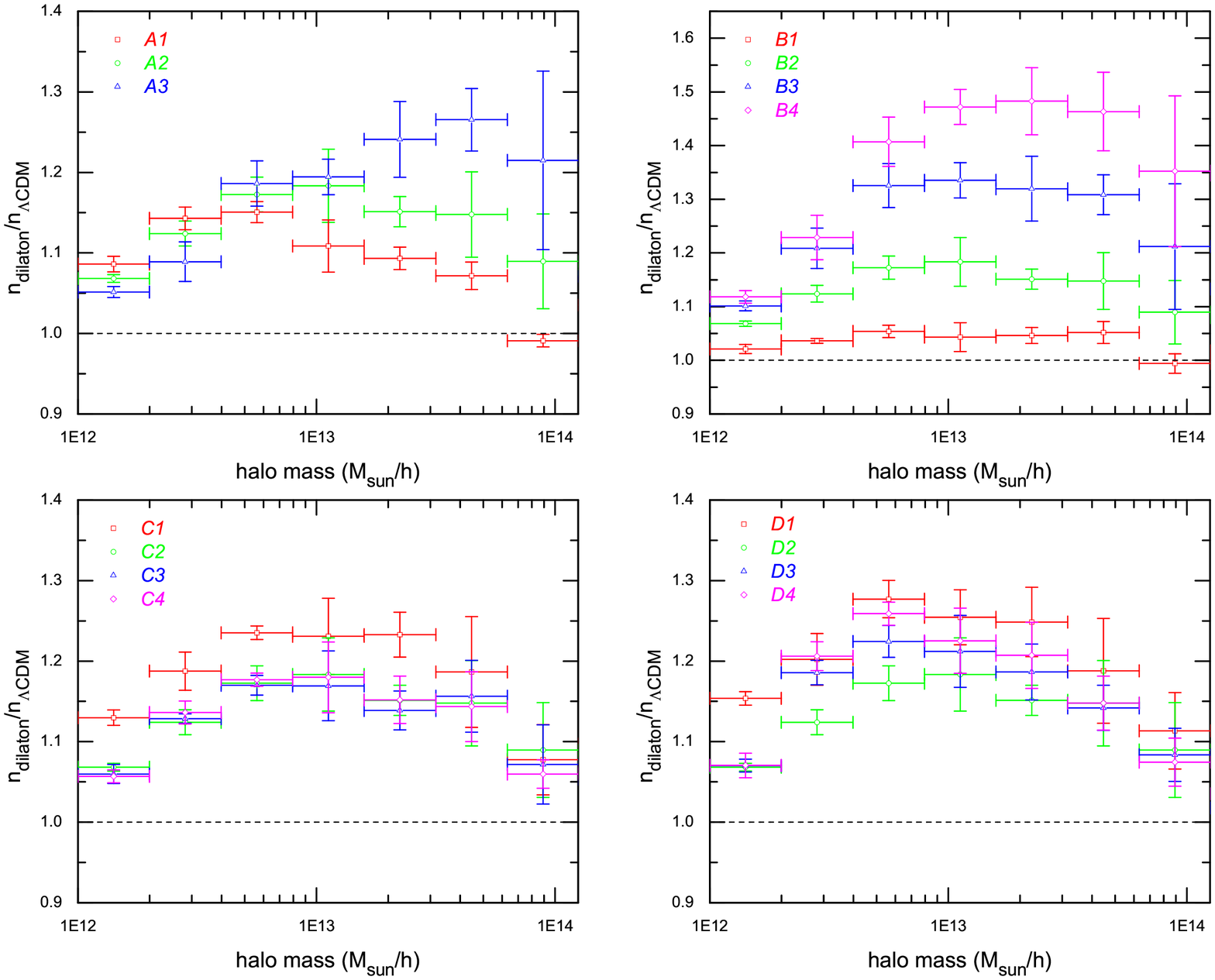}
\caption{(Colour online) The ratio between the mass functions of the dilaton models and the $\Lambda$CDM paradigm at $a=1.0$.} \label{fig:mf_z0}
\end{figure*}

\begin{figure*}
\includegraphics[scale=0.6]{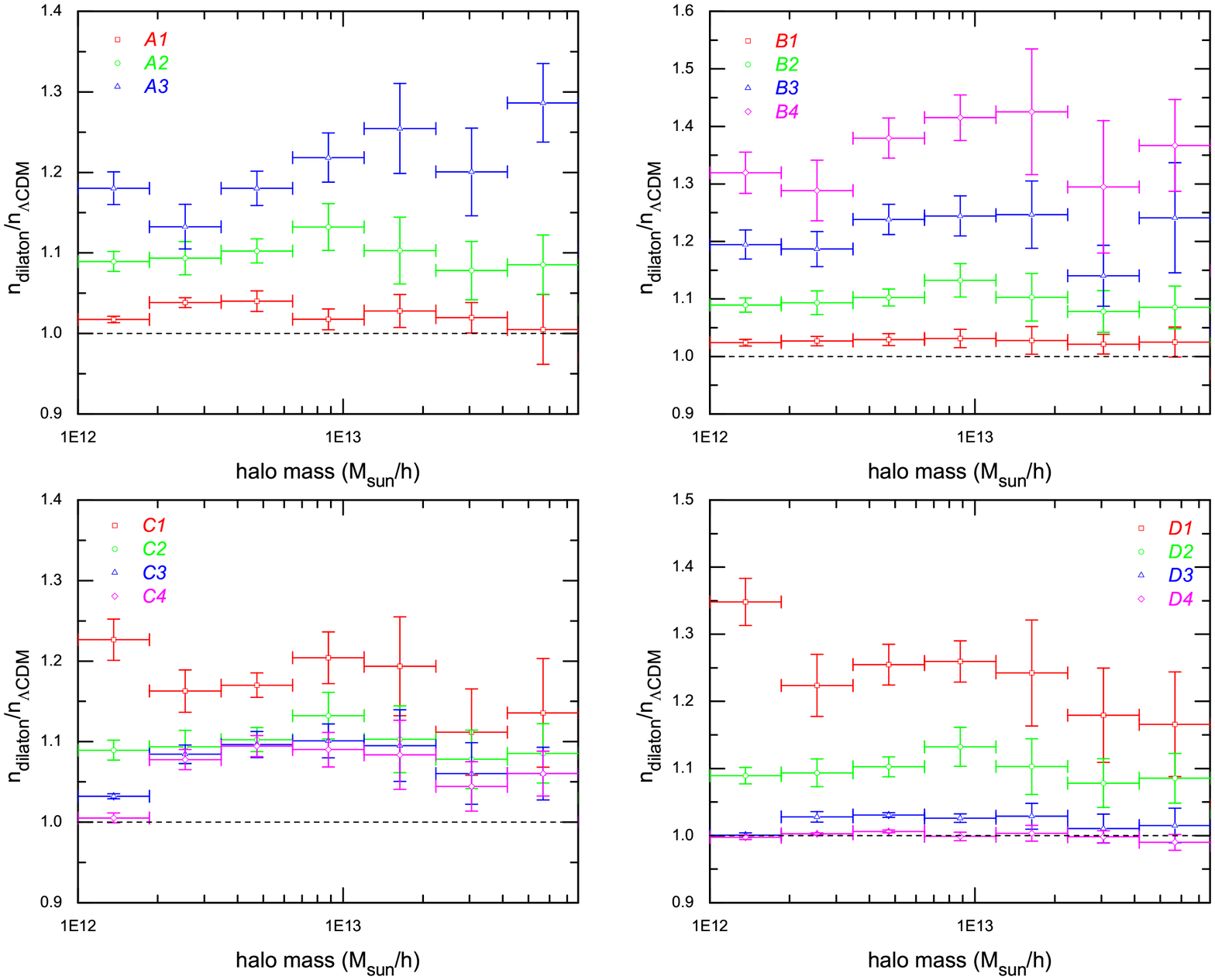}
\caption{(Colour online) The same as Fig.~\ref{fig:mf_z0}, but for $a=0.5$.} \label{fig:mf_z1}
\end{figure*}

\subsubsection{Nonlinear matter power spectra}

This subsection contains results about the nonlinear matter power spectra for the simulated dilaton models. \BLED{Fig.~\ref{fig:dpop_a_1.0} shows the relative differences between the dilaton and $\Lambda$CDM results at $a=1.0$, from which we can see the following properties:}
\begin{enumerate}
\item Decreasing $A_2$ leads to stronger matter clustering, since $A_2$ controls the steepness of the coupling function $A(\varphi)$ (see Fig.~\ref{fig:dilaton}). As discussed in \S~\ref{subsect:dilaton_effect}, the larger $A_2$ becomes, the steeper $A(\varphi)$ is and the harder it is for $\varphi$ to roll away from $\varphi_\ast$ where $\beta(\varphi)=0$ -- this means that $\beta$ is closer to zero and the fifth force is more strongly suppressed.
\item Increasing $\beta_0$ leads to stronger matter clustering, as $\beta_0$ determines the strength of the fifth force.
\item The $r$-dependence is weak since large changes in $\beta$ only take place at early times (see below). We see the feature discussed in \S~\ref{subsect:dilaton_effect}, that increasing $r$ decreases the matter power on larger scales ($k<0.2$Mpc/h) and increases it on smaller scales; this happens in both linear and nonlinear results.
\item As discussed in \S~\ref{subsect:dilaton_effect}, decreasing $\xi$ simultaneously {\it increases the strength} and {\it decreases the range} of the fifth force, causing more (less) clustering of matter on small (large) scales. This can be seen by comparing the results of D1 and D2. On even smaller scales, however, the matter power spectrum increases with $\xi$ again.
\item As in the symmetron case, at late times the linear perturbation theory is a rather bad approximation to the full nonlinear dilaton model, and fails to accurately predict the matter power spectrum even for $k\sim0.04h/$Mpc. This once again shows the important role $N$-body simulations have to play in the studies of modified gravity theories.
\item Overall, we see that the nonlinearity suppresses the matter power compared with the linear theory predictions, which shows that the dilaton mechanism works well for large scale structures. The suppression of the fifth force comes from two parts: the smallness of $\varphi$ and therefore $\nabla\varphi$ in high density regions, and the smallness of $\beta(\varphi)$ -- this indicates that with the same configuration of $\varphi$ the fifth force in the dilaton models here is more strongly suppressed than in the case of a constant $\beta(\varphi)$ \BLED{(e.g., in $f(R)$ gravity models), for which only the first part contributes to the screening}.
\end{enumerate}

At $a=0.5$ (cf.~Fig.~\ref{fig:dpop_a_0.5}), all the above properties remain, with the following noticeable features:
\begin{enumerate}
\item The agreement between linear perturbation theory and the full simulations gets better as nonlinearities have not reached their full effect. This is the same as the symmetron (see above) and $f(R)$ \citep{lhkzjb2012} cases.
\item The difference between the different C models becomes larger than at $a=1.0$ because, as mentioned above, the effect of changing $r$ is mainly to modify $\beta(a)$ at early times.
\end{enumerate}

The linear-nonlinear agreement is even better at $a=0.3$ (see Fig.~\ref{fig:dpop_a_0.3}). This indicates that the nonlinearity of the model only becomes important at late times, which is possibly because the formation of high density structures only then drives $\varphi$ to deviate from its background value.

Most of our simulation results show less deviation between the simulated dilaton models and $\Lambda$CDM than the case of the symmetron models. One of the reasons for this lies in the simulation details. In the symmetron models we have fixed the coupling strength $\beta_\star=1$, while for the dilaton cases, except for models B3 and B4, the coupling strength is taken to be at most $\beta_0\leq0.5$. As the fifth force scales as $\beta^2$, this makes a significant difference (c.f.~Fig.~\ref{fig:dpop_a_1.0}, upper right panel). As an example, model B4 differs from $\Lambda$CDM by nearly as much as the symmetron models do (and even more). 

\BLED{The shapes of the dilaton matter power spectra are worth discussing, as they show significant difference from the cases of symmetron and $f(R)$ gravity models. From Figs.~\ref{fig:dpop_a_1.0}, \ref{fig:dpop_a_0.5} and \ref{fig:dpop_a_0.3} we can see that:
\begin{enumerate}
\item In both linear and nonlinear cases, $\Delta P/P$ tends to flatten on small scales. In the linear case, this is very different from the behaviour of chameleon models with constant coupling strength $\beta$. In that case, the fifth force always has the same strength but at early times its range is limited by the very heavy scalar field mass: this means that on very small scales the fifth force has started enhancing clustering of matter ever since very early times, which is why $\Delta P/P$ keeps increasing with $k$ \citep{bdlw2012}. For dilaton models, on the other hand, the scalar field mass evolves more slowly and the coupling strength is suppressed at early times: this means that by the time the fifth force becomes non-negligible, its range has become large enough and below this range the growth of matter density perturbations is enhanced in a nearly scale-independent way (at least in the linear regime). Such a feature can indeed also be seen in the linear predictions of $\Delta P/P$ for symmetron models (cf.~Fig.~\ref{fig:dpop_a_1.0_sym}).
\item The flattening effect of $\Delta P/P$ on small scales is preserved when varying model parameters $A_2$ and $\beta_0$, but is weakened by varying $r$ and $\xi$. This is because, as discussed in \S~\ref{subsect:dilaton_effect}, varying $A_2$ and $\beta_0$ does not change the scalar field mass $m$, while varying the other two parameters does. Taking the parameter $r$ as an example, increasing $r$ makes $m$ more sensitively dependent on local matter density (i.e., more like a chameleon model which has no flattening in $\Delta P/P$). 	On the other hand, decreasing $r$ makes $\beta$ more sensitively dependent on local matter density and so suppresses the fifth force on large scales; on small scales the suppression can be compensated by the decreases of $m$, which makes $e^{-mr}$ larger, and the combined effect can be a weakened flattening of $\Delta P/P$ again.
\item Changes in $r$ (and similarly in $\xi$) make either $m$ or $\beta$ more sensitively dependent on local matter density, the deviation from linear perturbation results and the screening effect get stronger, especially at late times when structures have developed. This explains why at late times $\Delta P/P$ can decrease with time when varying $r$ and $\xi$.
\end{enumerate}
}

\BLED{The above results imply that the shape of the nonlinear matter power spectra can be different in dilaton and other modified gravity (e.g. chameleon) models. This will be studied in more details in a forthcoming work.}

\subsubsection{Mass functions}

This subsection contains the result of the mass functions from the dilaton simulations. The method to calculate the averages and standard deviations here is the same as that used in the symmetron case.

Fig.~\ref{fig:mf_z0} shows the results at $a=1.0$, where we can see that
\begin{enumerate}
\item The dilatonic fifth force enhances the formation of dark matter structures. The effect is strongest for medium-sized halos and is weaker for very large and very small halos. As in the symmetron case, this is because for very large halos the screening effect weakens this enhancement, and many of the small halos have accreted more matter or merged with other halos to form larger halos.
\item As discussed in \S~\ref{subsect:dilaton_effect}, decreasing $A_2$ makes the fifth force less screened, and as a result more large halos are formed \BLED{and fewer small halos survive the mergers}.
\item Increasing $\beta_0$ makes the fifth force stronger and produces more halos of {\it all} mass ranges probed by our simulations. \BLED{The dependence on $\beta_0$ is quite sensitive, for example, for $\beta_0=1$ the deviation from $\Lambda$CDM can be up to $~50$\%, while for $\beta_0=0.25$ this is less than 5\%.}
\item As in the case of the matter power spectrum, the mass function becomes larger as $r$ increases, and the dependence on $r$ is quite weak, especially when $r\leq1$ (models C2, C3 and C4). \BLED{As mentioned above, this is because increasing $r$ simultaneously increases the coupling strength and decreases the range of the fifth force, and the two effects cancel to some extent.}
\item The $\xi$-dependence of the mass function shows a similar behaviour to that of the matter power spectrum. For halos more massive than $\sim5\times10^{13}h^{-1}M_\odot$, we find that decreasing $\xi$ results in more halos being produced, similarly to the matter clustering power at $k\sim1h$Mpc$^{-1}$. For smaller halos, model D2 predicts fewest while D3, D4 gradually catch up D1, which is similar to the matter power at $k>3-4h$Mpc$^{-1}$. Overall, the $\xi$-dependence is quite weak, similar to the $r$-dependence.
\end{enumerate}

\BLED{As in the case of the matter power spectra, we are interested in the shapes of the mass functions. As discussed above, changing $r$ (or $\xi$) makes either the scalar field mass or the coupling strength more sensitively depend on local matter density, and in both cases the screening gets stronger (especially for large halos), consistent with what is seen in the matter power spectrum. A change in $A_2$ strengthens or weakens the screening effect but does not change the coupling strength for unscreened particles, and as a result the mass function behaves  as in $f(R)$ gravity models \citep{zlk2011}. Finally, a change in $\beta_0$ mainly affects the coupling strength for unscreened particles, but not so much the degree of screening, which is why $\Delta n/n$ flattens for large halo masses.}

To see how the dilaton effect on the mass function changes with time, we also show in Fig.~\ref{fig:mf_z1} the ratio between the mass functions at $a=0.5$. As discussed in the previous subsection, at this time the linear perturbation theory is a better approximation to the full theory. This implies that the screening of the fifth force has not yet been very significant, as is confirmed by this figure, which shows a weaker suppression of the dilaton-to-$\Lambda$CDM ratio at the high mass end. As in Fig.~\ref{fig:mf_z0}, the mass function results at $a=0.5$ show a good match with the behaviour of the matter power. \BLED{Note also that the effect of varying $r$ and $\xi$ is larger at early times, which also agrees with the behaviour of matter power spectra.}

The above results indicate that the period between $a=0.5$ and $a=1.0$ is an important era for the dilaton model, during which the structure formation is significantly affected by the nonlinearity of the model. In particular, we see that the shape of $\Delta P/P$ and $\Delta n/n$ experiences qualitative changes during this period.


\section{Discussions, Summary and Conclusions}

\label{sect:summary}

\subsection{Symmetron and dilaton screening}

\label{subsect:temp}

Modified gravity models vary according to their screening mechanisms by which the fifth force is suppressed in local environments. The Vainshtein mechanism works in theories of the Galileon type where a scalar field with non-canonical kinetic terms couples to matter in a reduced fashion in dense environments. Chameleons have an environment-dependent mass that becomes large enough to Yukawa suppress the fifth force in dense regions. Finally, the symmetron and the dilaton share a similar mechanism whereby the coupling of the scalar field to matter is field-dependent and can vanish in the presence of dense matter. What distinguishes these two types of models is their scalar potentials: a Mexican-hat for symmetrons and a monotonic function for dilatons. The coupling function for both types of models is a quadratic function\footnote{\BLED{Of course, other types of coupling functions can be used, as we have done  in the generalised symmetron model.}}.

Following the idea of \cite{bdl2011,bdlw2012}, the generalised dilaton and symmetron models studied here are completely specified by two temporal functions $m(a)$ and $\beta(a)$. These give the most general models with a quadratic coupling to matter and scalar field mass that is a power-law function of $a$ in the background cosmology for the generalised dilatons.  For the generalised symmetron models, the scalar field mass vanishes for $a\leq a_\ast$ and increases to its present cosmological value from then. In both models, the screening of the fifth force is achieved in high density regions where the scalar field is trapped near the minimum of $A(\varphi)$.
Yet the temporal dependences of the coupling to matter are drastically different: for generalised symmetrons it varies smoothly from a vanishing value for $a\le a_ \ast$ to its present value whereas the generalised dilatons it grows exponentially fast in the recent past of the Universe to reach its present value.

As discussed in \cite{bdlw2012}, the background expansion rate of such models is practically indistinguishable from that of the standard $\Lambda$CDM paradigm, so that the cosmological effects of the fifth force could only be seen in the large-scale structures. In this work, we have performed large-scale $N$-body simulations for the generalised dilatons and symmetrons, investigating in detail the effects of varying the dilaton and symmetron parameters on the nonlinear structures of the Universe. Some of these parameters are associated with the coupling to matter $\beta_0$ (\BLED{$\beta_\star$ for the symmetron case}), and $\xi$ which specifies the range of the fifth force on the cosmological background. A few extra parameters are used in the parameterisation to define the shapes of the potential and coupling function as  functions of the scalar field. For the dilatons, these parameters are $A_2,r$ and for the symmetrons they are $a_\ast$, $N$ and $M$.

Let us first discuss the common features of these models:
\begin{enumerate}
\item The coupling to matter $\beta_0$ (\BLED{or $\beta_\star$}) determines the overall strength of the fifth forces, and increasing them leads to more structures.
\item Decreasing $\xi$ leads to a shorter range for the fifth force and therefore a smaller enhancement of matter clustering\footnote{\BLED{In the dilaton case, changing $\xi$ also affects the coupling strength, making the dependence on $\xi$ more complicated.}}.
\end{enumerate}


\BLED{In the end, the effects on structure formation are mainly determined by how fast the fifth force evolves and how efficient it is screened in dense regions. An intuitive way to see this is to look at the expressions of $\beta(a)$ in these two models, as our discussion on tomography shows that this could be translated into $\beta(\rho_m)$, therefore giving us a sense about the screening, at least qualitatively. From Eqs.~(\ref{eq:beta_new_symmetron}, \ref{eq:beta_new_dilaton}) we can see that
\begin{enumerate}
\item In symmetron models, the coupling vanishes at $a\leq a_\ast$ (or equivalently for $\rho\geq\rho_\ast$) and after that it grows as a power-law function. Varying from $0$ to $\beta_\star$ between $a=a_\ast$ and today, $\beta$ depends quite sensitively on $a$ or $\rho_m$ in the regime with $\rho_m\leq\rho_\ast$; however, the symmetry of $V_{\rm eff}$ can be quickly restored for $\rho_m>\rho_\ast$ resulting in a strong suppression of the fifth force. In other words, there is a clear {\it cutoff} density beyond which the screening is very effective, and this cutoff is close to $\rho_\ast$, which is fairly low.
\item In dilaton models, the coupling grows exponentially \HANS{with time and with decreasing density}. As can be seen in Eq.~(\ref{eq:beta_new_dilaton}), $\beta$ decreases and becomes vanishingly small if one goes back in time or goes to high-density regions, much more quickly than it does in the symmetron models [c.f.~Eq.~(\ref{eq:beta_new_symmetron})]. This implies that the dilaton screening can become effective for lower densities than the symmetron mechanism.
\end{enumerate}
It appears  that the dilaton screening mechanism is more efficient than the symmetron mechanism. However, local tests of gravity are carried out in very dense regions, where the fifth force can be strongly suppressed in both models. Without specifying the exact parameter values for a given model, being it dilaton or symmetron, it is hard to say which one can satisfy local constraints more easily\footnote{\BLED{It is clear that by varying the parameter values both models can be made either more or less screened.}}.}



\subsection{Summary of numerical results}

Let us now summarise the results for each model.

\subsubsection{Generalised symmetron models}

The symmetron models we have simulated are close to what is allowed by local gravity experiments. Those constraints are mainly on the combination of the parameters $a_\ast$ and $\xi$ with the coupling strength $\BLED{\beta_\star}$ being an (almost) unconstrained parameter. This parameter, which controls the magnitude of the fifth force compared with gravity, can in principle be constrained by its effect on the cosmic structure formation.

Our simulations show that for a fiducial value of $\BLED{\beta_\star} = 1.0$ the symmetron models predict an enhancement of the nonlinear power spectrum with respect to $\Lambda$CDM of up to $40\%$ for $k \sim 1~h${Mpc}$^{-1}$ and up to $50\%$ at $k\sim 10~h$Mpc$^{-1}$. Likewise we find an enhancement of up to $50\%$ in the mass function for halo masses in the range of $10^{12}-10^{14}h^{-1}M_{\odot}$.

\BLED{We have shown how the fifth-force effect is changed by varying the other four model parameters: $a_\ast, N, M$ and $\xi$.}
\begin{enumerate}

\item \BLED{The parameter $a_\ast$ controls when the symmetry in $V_{\rm eff}(\varphi)$ is broken so the fifth force becomes non-vanishing. Decreasing $a_\ast$ gives it more time to influence the matter clustering, as a result not only the matter power spectra and mass functions deviate more from the $\Lambda$CDM results but also their shapes change qualitatively (more discussion below).}

\item \BLED{$N$ is the parameter which controls the coupling strength via $\beta\propto\varphi^N$. Since $|\varphi|$ is very small, increasing $N$ will suppress the magnitude of $\beta$ (or the fifth force), and therefore causes less clustering of matter.}

\item \BLED{$M$ is the {\it shape} parameter of the symmetron field potential, which determines how easy it is for $\varphi$ to roll away from $\varphi=0$ where $\beta$ vanishes. Increasing $M$ makes this easier, leading to a less-screened fifth force and thus more clustering and structures of matter.}

\item \BLED{$\xi$ controls the scalar field mass and therefore the range of the fifth force in vacuum, $\lambda_\star=2998\xi h^{-1}$Mpc. Increasing $\xi$ makes the scalar field mass (range of the fifth force) proportionally larger (shorter), and thus leads to a stronger suppression of the fifth force and limits its range.}
    
\end{enumerate}

As a rough guidance, increasing the symmetry-breaking scale factor $a_\ast$ from $0.33$ to $0.50$, decreasing $\lambda_\star$ from $2.0h^{-1}$Mpc to $1.0h^{-1}$Mpc, increasing $N$ from $2$ to $4$ or reducing $M$ from $6$ to $4$ are found to lower the enhancement of the power spectra and mass functions by $\sim10-20\%$. \BLED{The parameters we adopt in the simulations are in the `realistic' range and can be tested by future galaxy surveys.}

\subsubsection{Generalised dilaton models}

\BLED{We have also studied how structure formation in the generalised dilaton models is affected by varying the four model parameters $A_2, \beta_0, r$ and $\xi$.}
\begin{enumerate}
\item The effect of increasing $A_2$ is to make the total effective dilaton potential $V_{\rm eff}(\varphi)$ steeper and so to keep the scalar field closer to $\varphi_\ast$, where $\beta$ and the fifth force vanishes. The $\Lambda$CDM limit is retrieved by letting $A_2\rightarrow\infty$. According to our simulations, reducing $A_2$ to $5\times10^4$ produces a $\sim20\%$ enhancement in the nonlinear matter power spectrum between $z=1$ and $z=0$, which is significantly smaller than the linear perturbation predictions, demonstrating the efficiency of the dilaton screening mechanism. It also enhances the mass function by maximally $\sim25\%$ in the same redshifts. These numbers assume that $\beta_0=0.5$.
\item The effects of increasing $\beta_0$ are to strengthen the fifth force overall, and $\beta_0=0$ corresponds to the $\Lambda$CDM paradigm. The simulations show that even increasing $\beta_0$ to $1.0$ only causes $30-35\%$ enhancement in the matter power for scales smaller than $k\sim1h$Mpc$^{-1}$ between $z=1$ and $z=0$. This is at least $50\%$ smaller than the linear perturbation result, again showing that the fifth force is efficiently screened in dense regions. In the mean time, the mass functions are increased by up to $50\%$ with respect to the $\Lambda$CDM prediction. These numbers assume that $A_2=10^5$.
\item Increasing $r$ to $3/2$ simultaneously increases the strength and decreases the range of the fifth force. The $r$-dependence of the matter clustering is rather weak as a result of the cancellation due to these two opposite effects. Assuming $A_2=10^5$ and $\beta_0=0.5$, increasing $r$ to $1.333$ only enhances the matter power spectra by less than $10\%$ at $k\sim1h$Mpc$^{-1}$ and $15\%$ at $k\sim10h$Mpc$^{-1}$, which is again significantly smaller than the predictions of linear perturbation theory. The mass function  increases by up to $25\%$ in this case.
\item The effects of increasing $\xi$ are similar to those of decreasing $r$, and as a result the dependence on $\xi$ is also fairly weak.
\end{enumerate}

\BLED{Again, future galaxy surveys can place realistic constraints on the models studied here.}

\subsubsection{Highlights and comparisons}

In both the generalised symmeton and dilaton models, as in $f(R)$ gravity models \citep{lhkzjb2012},we find that at late times the linear perturbation theory fails to be a good approximation even for quite large scales ($k\sim0.05h$Mpc$^{-1}$). However, at earlier times it gives better agreement with the full simulations. This indicates that the environmental suppression of the fifth force becomes more important at late times when cosmic structures (very dense matter clumps) have already formed. This highlights the importance of numerical simulations in the study of (screened) modified gravity models.


\BLED{The deviations of matter power spectra and mass functions from $\Lambda$CDM in the symmetron and dilaton models are not directly comparable, because they depend on the exact parameter values used in each model. However, we can see that the shapes of $\Delta P/P$ and $\Delta n/n$ can be very different in the two models, which is probably a consequence of the different behaviour of the respective fifth forces.}

\BLED{At early times, $\Delta P/P$ increases with $k$ in both models (see e.g., Figs.~\ref{fig:dpop_a_0.5_sym} and \ref{fig:dpop_a_0.3}), similarly to what we see in $f(R)$ gravity models \citep{zlk2011,lhkzjb2012}. Differences appear at late time when the fifth force has been in effect for long enough:
\begin{enumerate}
\item For $f(R)$ gravity models we see that $\Delta P/P$ develops a peak at $k\sim\mathcal{O}(1)h$Mpc$^{-1}$, and on even smaller scales it decreases with $k$. The peak comes from the enhanced matter clustering due to the fifth force acting between clusters, and the turnover on small scales is because (compared with $\Lambda$CDM result) on these scales the short-range fifth force still accelerates particles and prevents them from further clustering\footnote{\BLED{Contrary to intuitive understandings, this is {\it not} because `the fifth force is suppressed on small scales'. The chameleon effect only reduces the range of the fifth force, but not its amplitude within that range.}}.
\item In the symmetron case, we also see the peak of $\Delta P/P$ at $k\sim\mathcal{O}(1)h$Mpc$^{-1}$, and on even smaller scales it goes up again. This seems to imply that the particle velocity inside halos stops being enhanced after the screening effect has kicked in (recall that the `cutoff' density for screening is quite low here and that `screening' here means a suppression of the amplitude, rather than range, of the fifth force), as a result of which the shape of $\Delta P/P$ on small scales is preserved since early times.
\item In the dilaton models, no obvious peak of $\Delta P/P$ can be seen: the power spectrum seems to have flattened on scales smaller than $k\sim1h$Mpc$^{-1}$. Such a flattening in $\Delta P/P$ is expected in the linear perturbation results for both the symmetron and dilaton models, as in the linear regime the time at which the fifth force becomes non-negligible is scale-independent below the scale $m_{0,\star}^{-1}$. For symmetrons the flattening is destroyed by the screening effect, while for dilatons it is not. As mentioned in \S~\ref{subsect:temp}, dilaton screening can apply  to lower matter densities: this indicates that the inter-cluster fifth force can be strongly suppressed as well, and thus the peak has not yet developed  (notice that in some cases, such as B4, there is a small bump). Again, a more definite conclusion could only be drawn after a more detailed study of the density and velocity fields in the simulations, which is beyond the scope of this paper.
\end{enumerate}
}

\BLED{The shape of $\Delta n/n$ at late times is similar in symmetron, dilaton and $f(R)$ gravity models, and the most important feature is that it goes down in the high-mass end, demonstrating efficient screening of the fifth force in these large structures. At early times, however, $\Delta n/n$ for the dilaton models show very weak mass dependence, which is close to the linear theory prediction, namely the fifth force is scale-independent.}

\BLED{In the symmetron models, varying the parameters $a_\ast, N, M$ and $\xi$ changes the shape of $\Delta P/P$ (and of $\Delta n/n$) in similar ways, which results in a degeneracy in these parameters. This is because all these parameters control the degree of screening of the fifth force.}

\BLED{This is not the case for the dilaton models, in which only a variation of $A_2$ changes the screening monotonically. Varying $\beta_0$ changes the overall strength of the fifth force more than its screening, while varying $r$ or $\xi$ changes the screening in more complicated ways. As a result there is no degeneracy in these parameters, except between $r$ and $\xi$ (see Fig.~\ref{fig:dpop_a_1.0}).}

\subsection{Conclusions and outlook}

\BLED{In short, the aim of this paper is threefold:
\begin{enumerate}
\item to show the power of the modified gravity parameterisation proposed in \citep{bdl2011,bdlw2012} in systematic studies of structure formation,
\item to acquire a sense about the qualitative behaviour of the generalised symmetron and dilaton models, and the effects of varying individual parameters, and
\item to make a preliminary exploration of the 4-dimensional parameter spaces in these models and find  models which are testable by the near-future observations.
\end{enumerate}
}

For all the test models in this paper, we find deviations from $\Lambda$CDM with similar magnitudes as those found in the $f(R)$ gravity model \cite{zlk2011,lhkzjb2012}, which means that many of the cosmological tests of $f(R)$ gravity \cite{lzk2012,zlk2011b,lzlk2012,jblzk2012} could in principle be carried out here as well.

On the other hand, the predictions of the cosmological observables can be different from those in other modified gravity models with screening mechanisms, such as the chameleon models. \BLED{For example, the shape of the matter power spectrum can be different in the symmetron, dilaton and $f(R)$ gravity models, which implies that the respective screening mechanisms indeed work quite differently. It would be interesting to understand better the origin of such differences and see if they can be used to distinguish between the different modified gravity models in cosmology. These studies are under way.}

\begin{acknowledgments}
ACD is supported in part by STFC. BL acknowledges supports by the Royal Astronomical Society and Durham University. HAW thanks the Research Council of Norway FRINAT grant 197251/V30 for support and Durham University for the hospitality where part of this work was carried out. GBZ is supported by STFC grant ST/H002774/1. The dilaton simulations were performed on the ICC Cosmology Machine, which is part of the DiRAC Facility jointly funded by STFC, the Large Facilities Capital Fund of BIS, and Durham University. The symmetron simulations were performed on the NOTUR Clusters TITAN, HEXAGON and STALLO, the computing facilities at the University of Oslo, Bergen and Troms{\o}.
\end{acknowledgments}

\appendix

\end{document}